\documentclass[12pt]{article}
\usepackage{fullpage}
\usepackage{amsmath}
\usepackage{amssymb}
\usepackage{amsthm}
\usepackage{graphicx}
\usepackage{fancyhdr}
\usepackage{feynmf}
\usepackage{cite}
\usepackage{hyperref}

\usepackage{color}
\definecolor{darkblue}{rgb}{0,0,0.5}

\begin{document}

\title{{\bf Jet Physics from the Ground Up}}
\author{Andrew J.~Larkoski\footnote{larkoski@slac.stanford.edu}\\ \\\small{Theory Group}\\\small{SLAC National Accelerator Lab}}
\date{}
\maketitle

\begin{abstract}
\noindent These are lecture notes presented at the online 2021 QUC Winter School on Energy Frontier hosted by the Korea Institute for Advanced Study.  They extend lectures presented at the 2017 and 2018 CTEQ summer schools in \cite{Larkoski:2017fip} and the 2020 Hadron Collider Physics Summer School hosted by Fermilab in \cite{Larkoski:2020jyz}.  Jets in quantum chromodynamics (QCD) are motivated from familiar results in classical electricity and magnetism, through identification of structures that exhibit approximate scale invariance.  From this point, an effective description of QCD jets is developed from simple assumptions that necessitate all-orders resummation for obtaining finite results.  With machine learning becoming an increasingly important tool of particle physics, I discuss its utility exclusively from the biased view for increasing human knowledge.  A simple argument that the likelihood for quark versus gluon discrimination is infrared and collinear safe is presented as an example of this approach.  End-of-lecture exercises are also provided.
\end{abstract}

\clearpage

\tableofcontents

\section{Lecture 1: Warm-Up: Jet Physics from Classical E\&M}

Hello!  I'm Andrew Larkoski, a professor at SLAC National Accelerator Laboratory, in Menlo Park, CA, USA, and I'm excited to lecture at this year's KIAS Winter School.  I have been tasked with discussing hadronic jets, which are collimated streams of particles created from the dynamics of quantum chromodynamics (QCD) at high energies.  In these lectures, I'll introduce jets from a ``bottom-up'' perspective, forgoing any explicit discussion of the fundamental QCD Lagrangian from which jets arise as an emergent phenomena.  This will be similar in approach to the recommended reading of my lectures from the CTEQ summer schools from a few years ago, Ref.~\cite{Larkoski:2017fip}, and my Hadron Collider Physics Summer School lectures, Ref.~\cite{Larkoski:2020jyz}.  Some of the philosophy of and calculations on jets presented here are also presented in my textbook on particle physics, Ref.~\cite{Larkoski:2019jnv}.  In this school, there will also be lectures on machine learning by Brian Ostdiek.  Especially in jet physics, machine learning has become an extremely widely-used tool for data analysis.  Now, as a theoretical physicist with research interests in calculations, I won't be discussing neural network architecture, programming, computer science, etc., but will instead propose the following.  Regardless of your individual thoughts of machine learning, the manner in which data is input and output of a machine suggests a novel way of thinking about fundamental problems in particle physics generally, and jet physics specifically.  I'll discuss this way of thinking and how it can be used to derive and learn about very general results in the final lecture.

\begin{figure}
\begin{center}
\includegraphics[width=5.5cm]{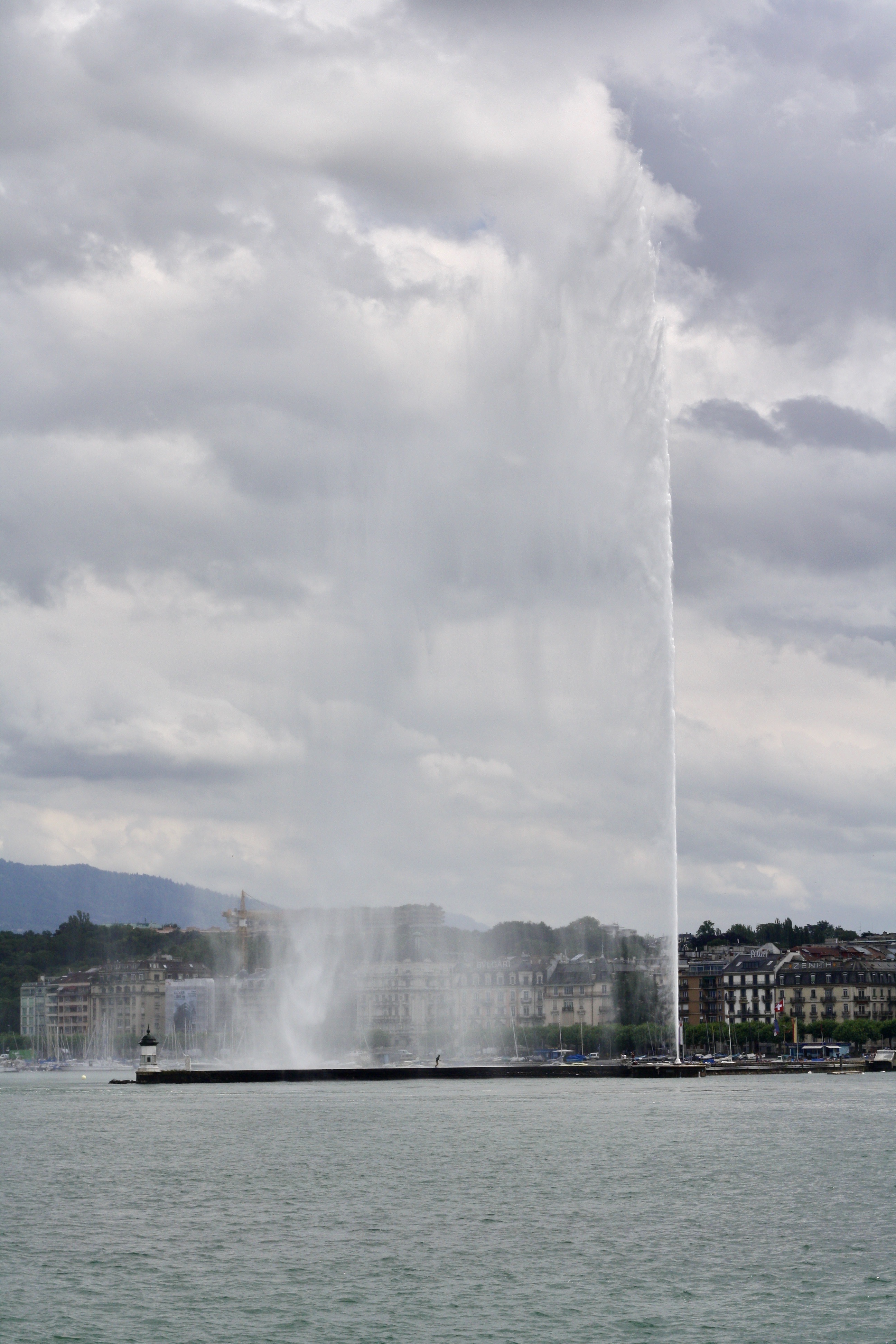}
\caption{\label{fig:jetdeau}
Photo of the Jet d'Eau located in Lake Geneva.
}
\end{center}
\end{figure}

As a first introduction, the word ``jet'' likely brings to mind exciting, fast things.  A jet airplane, for example, travels fast through the air, and while the jets of QCD will be traveling fast (like at or near the speed of light), this airplane image is not quite the right analogy.  A QCD jet is more similar to a jet of water, like that that comes out of a faucet or hose.  The water from the hose is traveling at high speed, but it is also relatively focused, or collimated, as it leaves the hose.  This enables you, for example, to spray off the dirt on that one corner of your car with precision.  In fact, one possible etymology of the particle physics word ``jet'' is from the Jet d'Eau, a spectacular 140 m water fountain that shoots out of Lake Geneva, near the city of Geneva, Switzerland.  The Jet d'Eau dates from the late 19th century, so has been a symbol of Geneva for well over a century, and has been associated with the science that occurs at nearby CERN for nearly 60 years.  A photo I took of the Jet d'Eau is shown in Fig.~\ref{fig:jetdeau}.

For this lecture, however, we will start very elementary, and introduce the notion of a jet from something with which you are already intimately familiar.  Let's go back to electromagnetism and think about, reinterpret, and challenge ideas that you were introduced to as an undergraduate.  To do this, let's just consider the properties of a point charge, which, for concreteness, we will take to be an electron:
\begin{center}
\includegraphics[width=0.7cm]{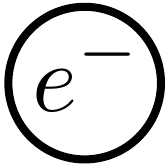}
\end{center}
With artistic license, I have drawn the electron as a sphere and with a label, but remember that it is a point, and has no spatial extent.  Now, if the electron just sits there motionless, it has an electric field, but that's about it:
\begin{center}
\includegraphics[width=3.5cm]{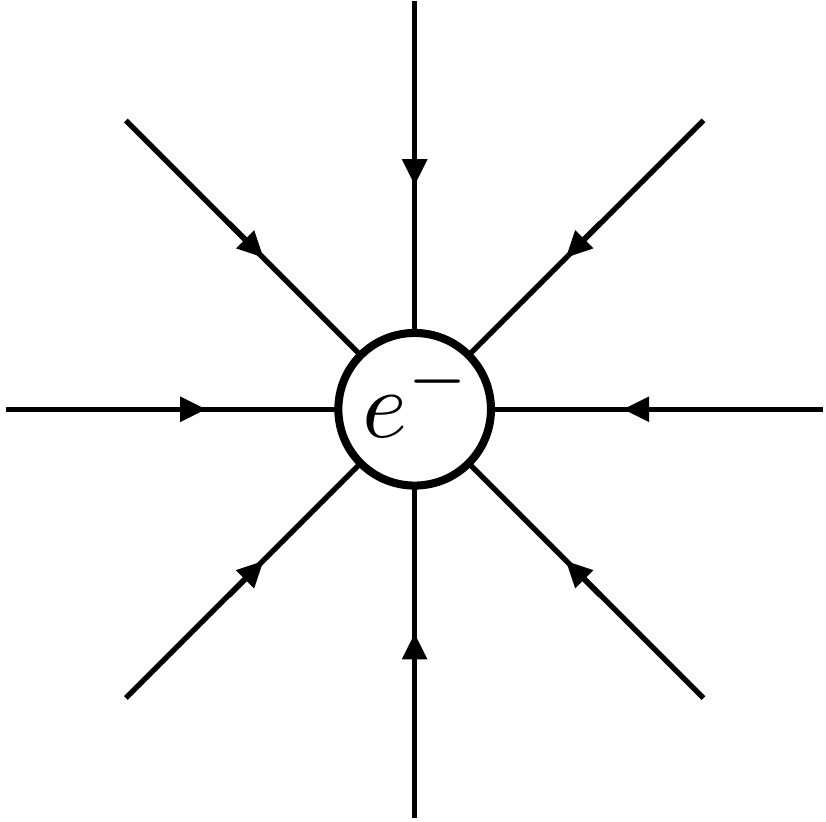}
\end{center}
While the electric field permeates space, it does not change in time and carries away no energy of the electron.  The energy contained in a particle's electric field is really just another way to represent the energy of the particle for simply existing.

In a study of classical electromagnetism, you learned that things get much more exciting if the charges move.  For instance, if the electron moves at a constant velocity, then not only does it have an electric field, but also a magnetic field.  What we see as an electric or magnetic field depends on our frame of reference with respect to the charged particle.  So, whenever we are studying electric charges at relativistic speeds, notions of ``electric'' or ``magnetic'' fields in isolation cease to have meaning.  There is just an electromagnetic field created by the particle whose dynamics we would like to understand.  We will be focusing our attention to relativistic particles, so our goal here will be to determine properties of the electromagnetic field that emanates from the electron.  Still traveling at constant velocity, the electromagnetic field of the electron is just a Lorentz transformation of the electric field of the particle at rest.  The field travels along with the electron, but sticks with the electron, and doesn't travel off on its own.  

\begin{figure}
\begin{center}
\includegraphics[width=5.5cm]{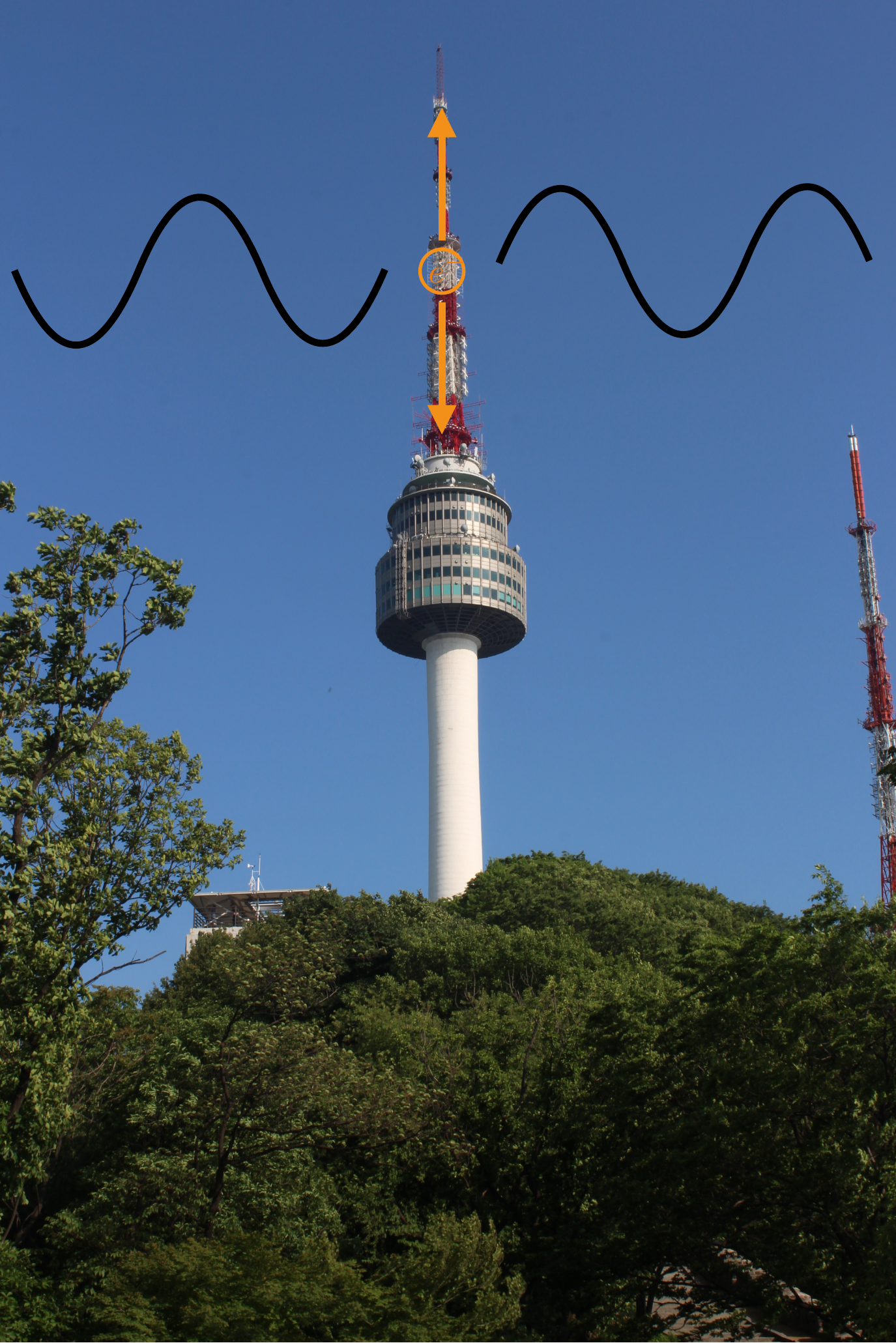}
\caption{\label{fig:namsan}
Photo of N Seoul Tower on Namsan with an added illustration of accelerating electrons in the antenna producing electromagnetic radiation.
}
\end{center}
\end{figure}

Of course, the whole story changes when the electron accelerates.  Accelerating charges correspond to acceleration of their dipole moment, which correspondingly induces electromagnetic radiation that carries energy away from the charges.  This property of classical electromagnetism is exploited to great effect in an antenna, like the N Seoul Tower at the top of Namsan.  Electrons are accelerated back and forth along a linear antenna, say, from some alternating current source.  These accelerating electrons then emit electromagnetic waves that travel at the speed of light away from the antenna, as illustrated in a photo I took on Namsan in Fig.~\ref{fig:namsan}.  The waves can then can be subsequently detected by another antenna, through essentially the opposite reaction: the electromagnetic radiation accelerates the charges in the antenna that correspondingly produce an alternating current and that can be observed on an oscilloscope.

Something important to note, however, is that the direction of the electromagnetic radiation from this accelerating electron is rather isotropic.  If the acceleration of the electron is relatively small, the electromagnetic radiation is not focused along any direction and is nothing like what could be called a ``jet''.  This is an extremely important property for a radio, for example, to ensure that wherever you are in Seoul you can listen to your music.

To emit a focused beam of electromagnetic radiation, we need extreme acceleration of our electron, and an antenna just isn't going to cut it.  Extreme accelerations are very easily accomplished in collisions, not unlike that you explore with carts in introductory physics, but here we will be colliding electrons.  Consider the collision of two electrons traveling at high speeds that bounce off one another, as illustrated below:
\begin{center}
\includegraphics[width=5cm]{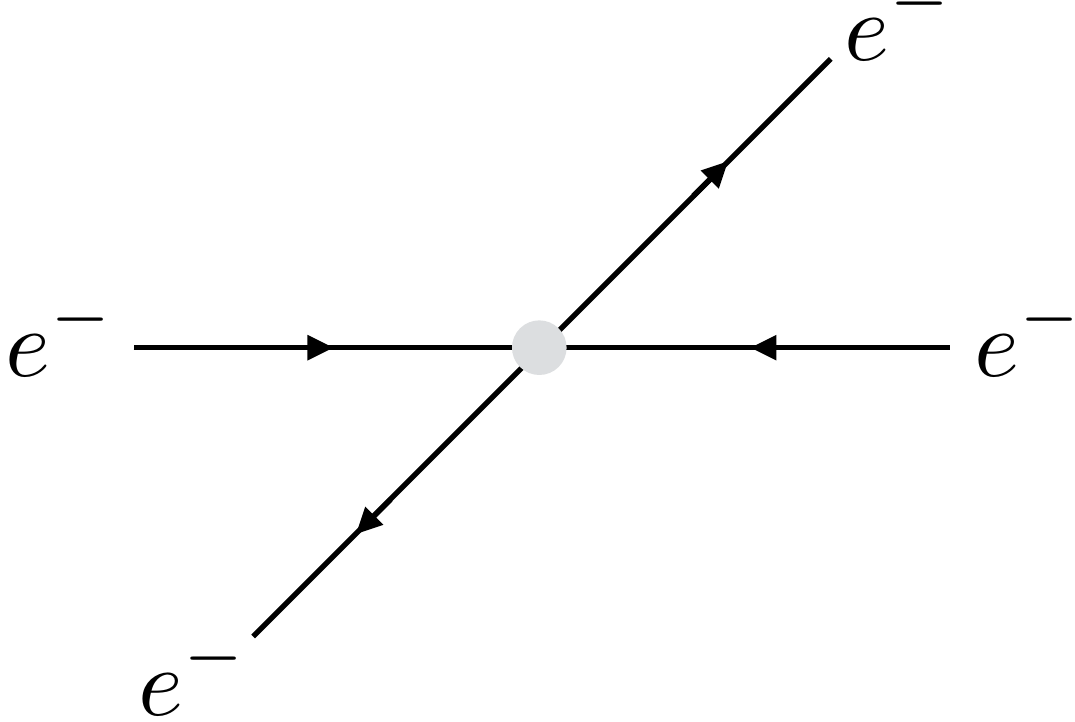}
\end{center} 
(I have obscured the region where the electrons get very close to one another.) With sufficiently high initial velocities, the electrons can get extremely close to one another before repelled by their electric fields.  Because of the inverse square law of the electric force, the region of strongest acceleration is isolated to their point of closest approach.  This point of closest approach can be estimated and at the very least illustrate the scales that one is considering with such a process.   To estimate the closest approach, we can use energy conservation.  Let's assume that the electrons were sent off to collide head-on from infinity with identical speed, and so their total relativistic kinetic energy would be
\begin{align}
K_i = 2(\gamma-1) m_e c^2\,,
\end{align}
where $\gamma$ is the boost factor and $m_e$ is the mass of the electron.   At their point of closest approach, the electrons will be (instantaneously) at rest and so their kinetic energy will have been transformed into electric potential energy:
\begin{align}
U_f = \frac{1}{4\pi\epsilon_0}\frac{e^2}{r}\,,
\end{align}
where $e$ is the electric charge of the electron, and $r$ is their distance of closest approach.  Setting the initial kinetic and final potential energies equal, the distance of closest approach is\footnote{The electrons are accelerating during their entire journey to collision, and are therefore radiating and losing energy the entire time.  We're ignoring this complication for simplicity of the estimation, but is something we will have to address later.}
\begin{align}
r = \frac{e^2}{8\pi\epsilon_0 m_e c^2}\frac{1}{\gamma-1}\,.
\end{align}
We can plug in the values for the physical constants, and find that the distance to closest approach is approximately
\begin{align}
r = \frac{1.4\times 10^{-15}\text{ m}}{\gamma-1}\,.
\end{align}
That is, if the electrons are traveling with a velocity that is any appreciable fraction of the speed of light relativistically ($\gamma \gtrsim 2$), the electrons come within a femtometer of each other.  At such short distances, electric forces are exceptionally strong, producing extreme accelerations of the electrons.

These extreme accelerations mean that significant energy is radiated from the point of collision in electromagnetic waves.  The total energy or power radiated can be calculated using results like the Larmor formula, but for our purposes here we are more interested in the relative direction of the radiation with respect to the electrons.  For an electron with acceleration $a$ in the same direction of its motion, the electromagnetic power $P$ radiated per unit solid angle $d\Omega$ is
\begin{align}
\frac{dP}{d\Omega} = \frac{\mu_0 e^2 a^2}{16\pi^2 c}\frac{\sin^2\theta}{(1-\beta \cos\theta)^5}\,.
\end{align}
(This is calculated in Example 11.3 in David Griffiths's electrodynamics textbook \cite{grifem}).  Here, $\theta$ is the angle with respect to the direction of electron's motion, and $\beta = v/c$, the speed of the electron as a fraction of the speed of light.  As anticipated from our simple picture of accelerating charges in an antenna, if the electron is traveling at low speeds, $\beta \to 0$, then the power distribution is smooth and relatively weakly dependent on the polar angle $\theta$, manifesting near-isotropy of power in radiation.  However, for speeds close to that of light $\beta \sim 1$, the angle of maximum power radiation is very close to the direction of the electron's velocity:
\begin{align}
\theta_{\max} \simeq \sqrt{\frac{1-\beta}{2}} \,.
\end{align}
For speeds $\beta$ near 1, this can also be related to the electron's total energy $E$ and mass $m_e$, where
\begin{align}
1 -\beta \simeq \frac{m_e^2 c^4}{2E^2}\,.
\end{align}
Then, the angle of maximum power in electromagnetic radiation can be expressed as
\begin{align}
\theta_{\max} \simeq \sqrt{\frac{1-\beta}{2}} \simeq \frac{m_e c^2}{2E}\,.
\end{align}

This result is profound.  We can augment our simple picture of electron-electron scattering with the knowledge gained from this analysis.  For high electron energies, $E \gg m_ec^2$, there is a significant amount of electromagnetic radiation emitted at very small angle with respect to the electrons' velocities:
\begin{center}
\includegraphics[width=5cm]{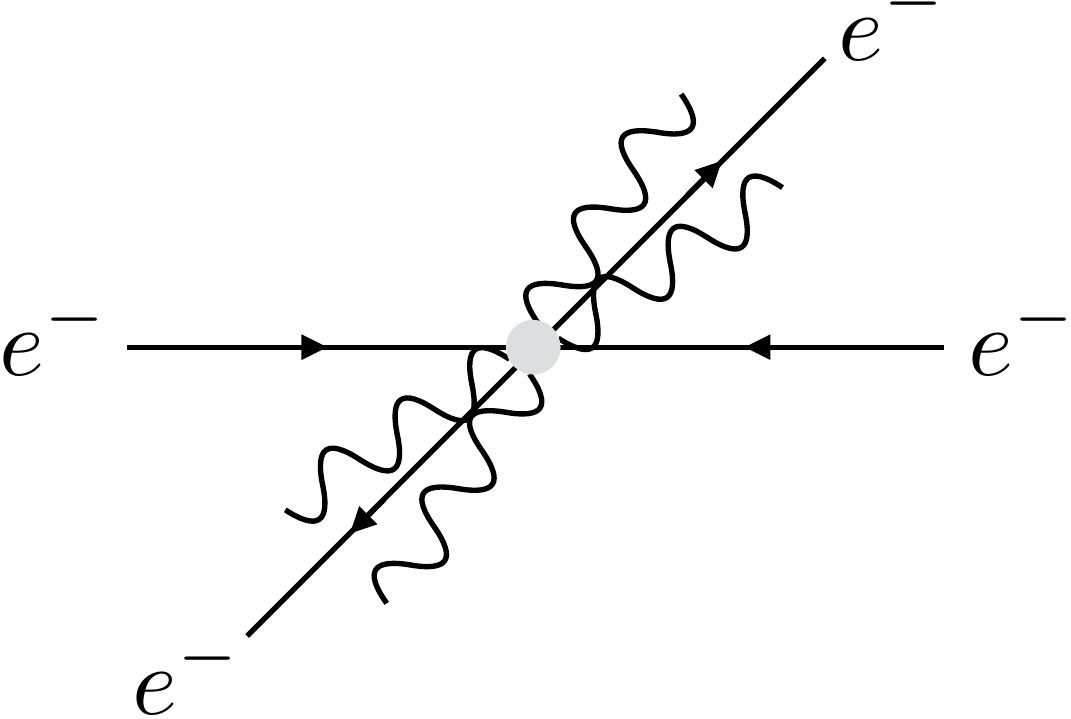}
\end{center} 
I have just drawn a few electromagnetic waves as the sinusoidal curves for illustration.  At asymptotic energies, or in the approximation that the electrons are massless, there is an unbounded amount of electromagnetic power radiated along the direction of motion of the electrons.  This so-called ``collinear divergence'' results in a narrow, bright beam of light in the direction of the electrons.  We could therefore refer to this high-energy electron and the electromagnetic radiation emitted from collision as an electron jet.

This isn't yet the complete story.  In classical mechanics, electromagnetic radiation is, of course, treated as a classical wave.  However, it is treated a bit magically compared to other classical waves with which we are familiar.  Electromagnetic waves can propagate in vacuum, while other waves require a medium, and there is apparently no substructure to electromagnetic waves.  They just {\it are}.  Addressing these confusions is clearly not necessary for establishing a huge array of quantitative predictions of electromagnetism, but it should leave a lot to be desired in terms of claiming complete understanding.  For concreteness, an electromagnetic wave seems (on the surface) to be vastly different than a water wave.  A wave of water is a coherent oscillation of water (the medium) by an enormous number of H$_2$O molecules.  ``Coherent'' means that the individual microscopic molecules are moving in a strongly-correlated fashion to exhibit a macroscopic wave.  A water wave is also called a ``collective phenomenon'' because a large collection of molecules (comparable or more molecules than Avogadro's number $N_A$) exhibit a phenomenon that is completely unapparent from the individual molecules' motion.

If you just look at waves on the ocean, for example, you don't observe individual water molecules; you see the wave as an object, undulating, breaking, and crashing on the beach.  It is due to the sheer enormity of Avogadro's number $N_A \simeq 6.02 \times 10^{23}$, that this is possible, and indeed why the reality of atoms was not established for hundreds of years after Newton's laws.  With this perspective, however, the confusions about electromagnetic radiation could potentially be solved.  If an electromagnetic wave were actually composed of an enormously large number of coherent particles, then we wouldn't need to worry about the necessity of a medium for the wave.  A free particle can happily travel along in a vacuum at a constant speed.  Additionally, electromagnetic radiation travels at the speed of light, $c$, as it is, well, light.  Therefore, the only way this is possible is if the particles that compose the wave are themselves massless.  Of course, this should be entirely familiar as photons are the particles of light of which electromagnetic waves are a manifestation of their collective phenomena.

This interpretation of electromagnetic radiation results in another physical feature present in our electron scattering.  Energy is conserved in the collision, and the initial energy of the electrons is fixed, so there is only a finite amount of energy that could possibly be radiated away in electromagnetic waves.  Let's call the total energy in electromagnetic radiation $E_\text{EM}$.  Just observing an electromagnetic {\it wave} means that there must have been an enormous number of photons $N_\gamma \ggg 1$ emitted from the collision.  Then, the mean energy per photon $\langle E_\gamma\rangle$ is
\begin{align}
\langle E_\gamma\rangle = \frac{E_\text{EM}}{N_\gamma}\,.
\end{align}
The energy in any given photon is then extremely tiny.  Further, because the energy in any one photon is so small, we can freely add one, two, three, or more photons without really affecting the total energy.

However, this is a bit of a dirty way to say that photon number is not conserved.  This should concern you, so let's argue from a different perspective.  We had said that photons must be massless because electromagnetic waves travel at the speed of light.  Therefore, the energy of an individual photon has no non-trivial lower bound.  As a massless particle, its energy is just set by its momentum $\vec p$ or frequency $f$:
\begin{align}
E = |\vec p|c = hf\,,
\end{align}
where $h$ is Planck's constant.  All that is required is that $f\geq 0$, but the energy of a photon can be arbitrarily small.  So, we can easily boost the number of photons in our wave, to make it more wave-like, while at the same time not affecting its total energy by adding 0 energy or 0 frequency photons.  We can clearly add an arbitrary number of 0 energy photons and there would be no observable consequences.  We therefore refer to this as a ``low energy'' or ``infrared'' divergence because an electromagnetic wave contains a divergent number of low energy (i.e., lower-than-red frequency) photons.  From a statistical mechanics perspective, we would say that photons have no chemical potential; no physical meaning can be ascribed to the statement ``the number of photons''.

We have therefore argued that electromagnetism with massless electrons and particle photons exhibits two divergences, which have physical consequences.  The collinear divergence means that electromagnetic radiation is emitted dominantly along the direction of electron motion, and the infrared divergence means that every electromagnetic wave consists of an arbitrary number of arbitrarily low-energy photons.  These two phenomena imply that there is a scale-invariance of this physical system.  By ``scale invariance'' I mean that we can multiply quantities by some number $\lambda > 0$ and we would still see the same physical phenomena.  For example, for a massless electron with $\beta = 1$, the power radiated per unit solid angle in the small angle $\theta \to 0$ limit reduces to
\begin{align}
\frac{dP}{d\Omega}=\frac{\mu_0 e^2 a^2}{16\pi^2 c}\frac{\sin^2\theta}{(1-\cos\theta)^5}\to \frac{2\mu_0 e^2 a^2}{\pi^2 c}\frac{1}{\theta^{8}}\,.
\end{align}
Scaling the angle as $\theta \to \lambda \theta$ keeps the functional form of the radiated power the same:
\begin{align}
\left.\frac{dP}{d\Omega}\right|_{\theta\to\lambda \theta} \to \lambda^{-8}\frac{2\mu_0 e^2 a^2}{\pi^2 c}\frac{1}{\theta^{8}}\,.
\end{align}
Additionally, we can freely scale the energy of 0 energy photons by any factor $\lambda > 0$, and their energy remains, well, 0.  In the next lecture, we will take this scale invariance property as an assumption and demonstrate that jets are an inevitable consequence.

\subsection{Exercises}

\begin{enumerate}
\item A coherent state $|\chi\rangle$ in quantum mechanics is an eigenstate of the lowering operator $\hat a$ of the harmonic oscillator:
\begin{align}
\hat a|\chi\rangle = \lambda |\chi\rangle\,,
\end{align}
where $\lambda \in \mathbb{C}$ is the eigenvalue.  Solve this eigenvalue equation for the state $|\chi\rangle$ in terms of the eigenvalue $\lambda$ and the ground state of the harmonic oscillator, $|\psi_0\rangle$.  Recall that the harmonic oscillator Hamiltonian is
\begin{align}
\hat H = \hbar \omega\left(
\hat a^\dagger \hat a + \frac{1}{2}
\right)\,.
\end{align}
The lowering $\hat a$ and raising $\hat a^\dagger = (\hat a)^\dagger$ operators satisfy the commutation relation
\begin{align}
[\hat a,\hat a^\dagger]  =1\,,
\end{align}
and the ground state is annihilated by the lowering operator:
\begin{align}
\hat a|\psi_0\rangle = 0\,.
\end{align}

Show that for $\lambda \neq 0$, the coherent state $|\chi\rangle$ has contributions from every energy eigenstate of the harmonic oscillator.

\item A coherent state of low energy photons in the quantum theory of electromagnetism, quantum electrodynamics (QED), is described by a so-called Wilson line \cite{Wilson:1974sk}.  They were also identified by Fadde'ev and Kulish for forming finite, non-divergent predictions in QED \cite{Kulish:1970ut}.  A Wilson line between two spacetime points $x$ and $y$, $W(x,y)$, satisfies the differential equation
\begin{align}
\left(
\frac{\partial}{\partial x^\mu} - ieA_\mu(x)
\right)W(x,y) = 0\,.
\end{align}
Here, $A_\mu(x)$ is the electromagnetic vector potential at spacetime position $x$.

\begin{enumerate}

\item Solve this differential equation for $W(x,y)$.  Compare it to the quantum mechanical coherent state.

\item {\it Challenge!} In QED, $A_\mu(x)$ creates and annihilates photons, from the mode expansion:
\begin{align}
A_\mu(x) = \int \frac{d^3p}{(2\pi)^3}\frac{1}{\sqrt{2E_{\vec p}}} \sum_{r = 0}^3\left(
\hat a_{\vec p}^r\, \,\epsilon_\mu^r(\vec p\,)\,e^{-ip\cdot x}
+\hat a_{\vec p}^{r\dagger}\,\epsilon_\mu^{r*}(\vec p\,)\,e^{ip\cdot x}
\right)\,.
\end{align} 
Here, $\vec p$ is the three-momentum of the photon, $r$ is its polarization label, $\epsilon_\mu^{r}(\vec p\,)$ is the polarization vector, and $\hat a_{\vec p}^{r}$ ($\hat a_{\vec p}^{r\dagger}$) is the photon annihilation (creation) operator.  Insert this mode expansion into the expression you found for the Wilson line.  How does it compare to the quantum mechanical coherent state?  Also, express the Wilson line in normal order, where all annihilation operators are to the right of creation operators.  You will need the commutation relation
\begin{align}
\left[
\hat a_{\vec p}^r\,,\hat a_{\vec p\,'}^{r'\dagger}
\right] = (2\pi)^3 \delta^{rr'}\delta^{(3)}(\vec p-\vec p\,')\,.
\end{align}

\end{enumerate}

\end{enumerate}

\clearpage

\section{Lecture 2: QCD as a Weakly-Coupled Conformal Field Theory}

For this lecture, let's understand what QCD is in the first place, motivated by the observations that we made in electromagnetism.  My introduction will be very different (likely) than you've seen before.  To define QCD, jets and their consequences, we will only make two assumptions or axioms from which everything in these lectures follows.  They are:
\begin{enumerate}

\item At high energies, QCD is an approximately scale-invariant quantum field theory.  This means that, to first approximation, the coupling of QCD, $\alpha_s$, is constant, independent of energy.

\item Not only is $\alpha_s$ (approximately) constant, but further $\alpha_s$ is small; formally we assume that $\alpha_s \ll 1$.  This means that the degrees of freedom in the Lagrangian of QCD, quarks and gluons, are good quasi-particles for actually describing the physics of QCD.

\end{enumerate}

These assumptions are sufficient to write down matrix elements for some simple processes.  Actually, we'll just write down the probability density functions, because ``matrix element'' (implicitly) assumes specific interactions and particle properties, but our axioms say nothing explicit about interactions.  By axiom 2, we can ask questions about the dynamics of quarks and gluons.  So, we'll ask: what is the probability for a quark to emit a gluon?  With a Lagrangian and Feynman rules, we would want to calculate the (squared) diagram:
\begin{equation}
P_{qg\leftarrow q} = \left|
\raisebox{-1.2cm}{\includegraphics[width=5cm]{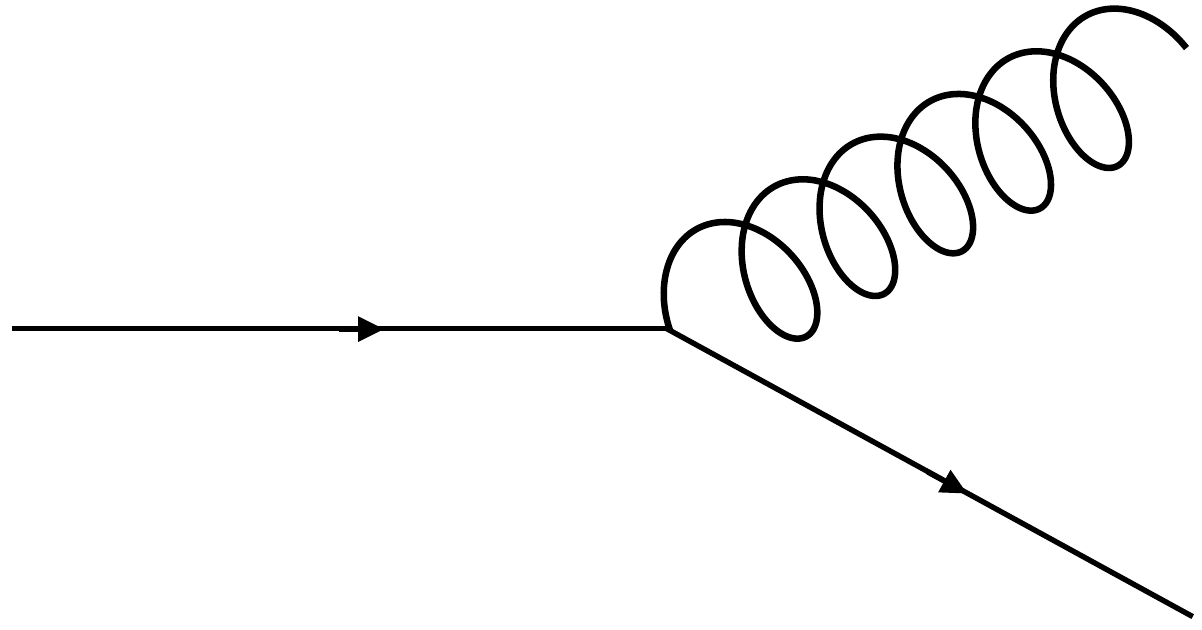}}
\right|^2\,.
\end{equation}
But we don't have Feynman rules, so we have to use our axioms.

We have a few things to establish before our axioms, however.  The probability for gluon emission will, in general, depend on the four-momentum of the gluon.  So, we need to identify the space on which the probability $P_{qg\leftarrow q}$ is defined.  Then, given that space, we can ask what constraints scale invariance imposes.  That is, we need to identify the degrees of freedom of the emitted gluon.

Let's start with the gluon's four-momentum written as:
\begin{equation}
p_g = (E,p_x,p_y,p_z)\,.
\end{equation}
The gluon is massless, so demanding that this momentum be on-shell requires $E = |\vec p|$, or that the momentum can be expressed in spherical coordinates as
\begin{equation}
p_g = E(1,\sin\theta \cos\phi,\sin\theta\sin\phi,\cos\theta)\,.
\end{equation}
Here, $\theta$ is the quark-gluon opening angle and $\phi$ is the azimuthal angle about the quark:

\begin{center}
\includegraphics[width=5cm]{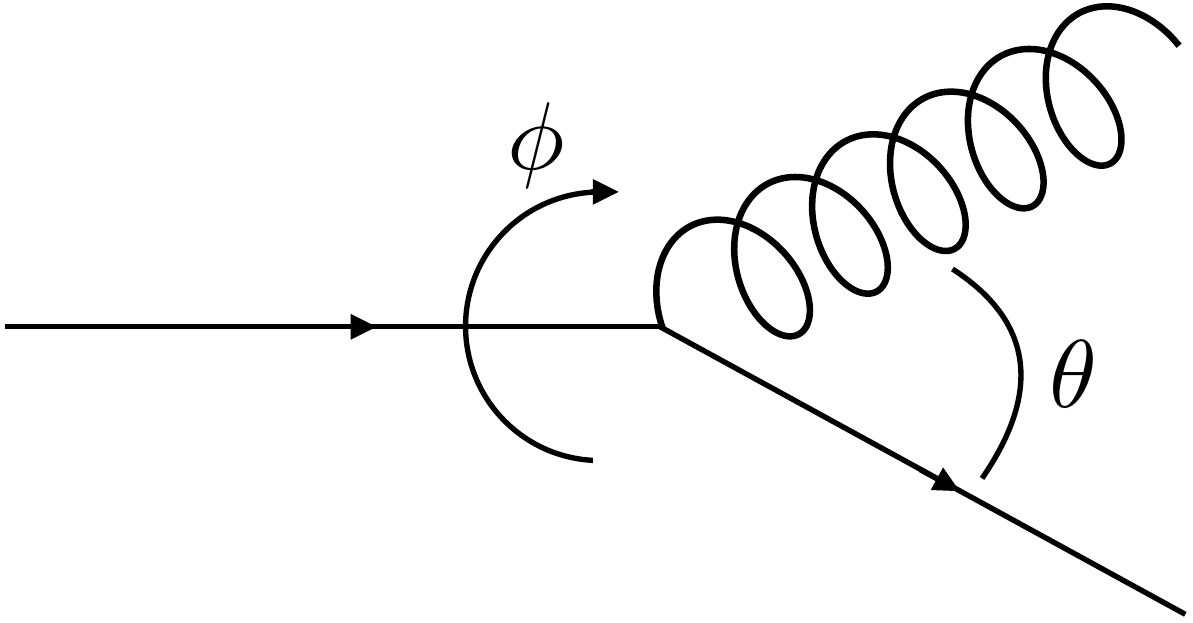}
\end{center}
Additionally, the total quark plus gluon energy is fixed.  Calling this total energy $E_\text{tot}$, we can write $E = zE_\text{tot}$, for some fraction  $z\in[0,1]$.  Additionally, from the formulation of our assumptions and the problem at hand, there was nothing special about the azimuthal angle $\phi$: the physics is independent of $\phi$.  Thus, we can fix $\phi$ to a convenient value, say, $\phi = 0$.  (That is, we assume that the quark is unpolarized.)  With these identifications, the gluon's momentum is:
\begin{equation}
p_g = zE_\text{tot}(1,\sin\theta,0,\cos\theta)\,.
\end{equation}
We had assumed that the final state quark's momentum is along the $z$-axis, so it is
\begin{equation}
p_q = (1-z)E_\text{tot}(1,0,0,1)\,.
\end{equation}
We have therefore identified the gluon's relevant degrees of freedom to be the energy fraction $z$ and angle $\theta$.  The splitting probability is:
\begin{equation}
P_{qg\leftarrow q} = p(z,\theta)\, dz\, d\theta\,,
\end{equation}
for some other function $p(z,\theta)$.

There are a few things we can immediately say about the function $p(z,\theta)$.  $\alpha_s$ is the QCD coupling; as such, it controls the strength with which quarks and gluons interact with one another.  Thus, this function, as it describes the probability of emission of a gluon off of a quark, is proportional to $\alpha_s$:
\begin{equation}
p(z,\theta) \propto \alpha_s\,.
\end{equation}
Actually knowing QCD we can say more, so we might as well add it.  The actual factors that come with $\alpha_s$ are:
\begin{equation}
p(z,\theta)\propto \frac{2\alpha_s}{\pi}C_F\,,
\end{equation}
where $C_F$ is the fundamental quadratic Casimir of the SU(3) color symmetry of QCD.  Again, our assumptions can't tell us about these factors, but they won't qualitatively change the picture we are developing (yet).  In QCD, $C_F = 4/3$, and so $2\,, \pi\,, C_F \sim {\cal O}(1)$, so truly $\alpha_s$ is what is controlling the coupling of quarks and gluons, by the assumption that $\alpha_s \ll 1$.  By the way, $C_F$ is a measure of how quarks and gluons share the three colors of QCD.

Next, we'd like to determine the dependence of $p(z,\theta)$ on the energy fraction $z$ and angle $\theta$.  To do this, we need to think about what our assumption of ``scale invariance of QCD'' means.  QCD is a quantum field theory, and as such is Lorentz invariant.  Thus, the only quantities that all observers agree on are those that are, well, Lorentz invariant.  If we further state that QCD is scale-invariant, this means that a scaling of all Lorentz-invariant quantities produces the same physics.  That is, probability distributions of Lorentz invariant quantities should be further invariant to scale transformations.

Given our quark-gluon system, the only Lorentz invariant we can construct is the dot product of their momenta:
\begin{equation}
p_q\cdot p_g = z(1-z)E^2_\text{tot}(1-\cos\theta)\,.
\end{equation}
This is Lorentz invariant by construction and scale invariance means that the scaling $p_q\cdot p_g \to \lambda\, p_q\cdot p_g$, for any $\lambda>0$, leads to identical physical phenomena.  In general, this scaling is not simply implemented on the energy fraction or angle, but there is a limit in which it is simple.  If the gluon has low energy or is soft so that $z\ll 1$ and is nearly collinear with the quark so that $\theta \ll 1$, note that
\begin{align}
&z(1-z)\xrightarrow{z\ll 1} z\,, &1-\cos\theta \xrightarrow{\theta\ll 1} \frac{\theta^2}{2}\,.
\end{align}
Then, in this double limit, the dot product is
\begin{equation}
p_q\cdot p_g \xrightarrow{z\ll 1\,, \theta\ll 1} z\theta^2\frac{E_\text{tot}^2}{2}\,.
\end{equation}
Thus, the soft and collinear limit corresponds to this dot product expressed as a power law function in both $z$ and $\theta$.  A scaling of $p_q\cdot p_g$ can therefore be accomplished by scaling of either $z$ or $\theta$ (or both).  Then, we identify the scale transformation under which QCD is invariant as
\begin{equation}
z\to \lambda\, z\qquad \text{or} \qquad \theta^2 \to \lambda \, \theta^2\,, \qquad \text{for }\lambda > 0\,.
\end{equation}

Before continuing, note that there is a coordinated scale transformation under which the product $z\theta^2$ is unchanged.  If we scale
\begin{equation}
z\to \lambda\, z\qquad \text{and} \qquad \theta^2 \to \frac{\theta^2}{\lambda}\,, 
\end{equation}
then $z\theta^2\to z\theta^2$.  That is, such a coordinated scale transformation is actually a Lorentz transformation, a boost long the direction of the quark's momentum.  Lorentz invariance states that if energies increase, angles must decrease to ensure that dot products are unchanged.

Now, if scale transformations are implemented by independent scalings $z\to\lambda_1\, z$ and $\theta^2 \to \lambda_2\, \theta^2$, for $\lambda_1,\lambda_2>0$ and the probability $P_{qg\leftarrow q}$ must be unchanged under this scaling, this uniquely fixes the $z$ and $\theta$ dependence of $p(z,\theta)$ to be:
\begin{equation}\label{eq:scalesplit}
P_{qg\leftarrow q} = p(z,\theta)\, dz\, d\theta = \frac{2\alpha_s}{\pi}C_F\, \frac{dz}{z}\frac{d\theta}{\theta}\,.
\end{equation}
This is invariant to the scalings and therefore satisfies our first assumption about QCD.  Note that for this simple functional form, we had to work in the soft and collinear limit.  While this may seem restrictive, we'll be able to get a lot of mileage out of it.

The first thing to note about this probability distribution is that, in the $z$ or $\theta$ to 0 limits, it diverges.  Actually, it's worse than that: not only does it diverge, but it is not even integrable as $z,\theta\to 0$!  Thus it isn't a ``probability distribution'' at all, because it cannot be normalized.  So how can such a distribution describe some physical process which is non-singular?  Note that the $z$ or $\theta$ to 0 limits are those limits in which the gluon becomes unobservable.  The $z\to 0$ limit is the limit in which the gluon has no energy. A detector, like the calorimeters at ATLAS or CMS, requires a finite energy of the particle to observe it: it never ``sees'' a 0 energy ``hit.''  The $\theta\to 0$ limit is when the gluon is collinear with the quark.  Because it is traveling in the exact same direction as the quark, there is no measurement you can perform to separate them out.  The angular resolution of the cells of the calorimetry at ATLAS and CMS is finite: the particles must be a non-zero angle from one another to be distinguished.

That is, there is no measurement you can perform in the $z\to 0$ or $\theta\to 0$ limit to distinguish a quark emitting one gluon to the case in which the quark emits no gluons.  That is, we call the $z\to 0$ or $\theta\to 0$ limits degenerate, as the physical configuration degenerates to a system with fewer gluons.  But this also points to an extension.  There is no measurement we can perform to distinguish a quark that emitted no gluons:

\begin{center}
\includegraphics[width=3cm]{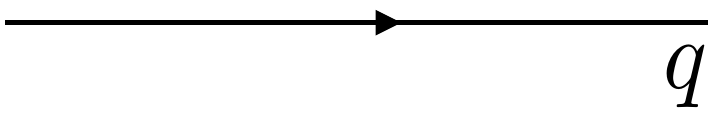}
\end{center}
from a quark that emitted one soft and/or collinear gluon:

\begin{center}
\includegraphics[width=3cm]{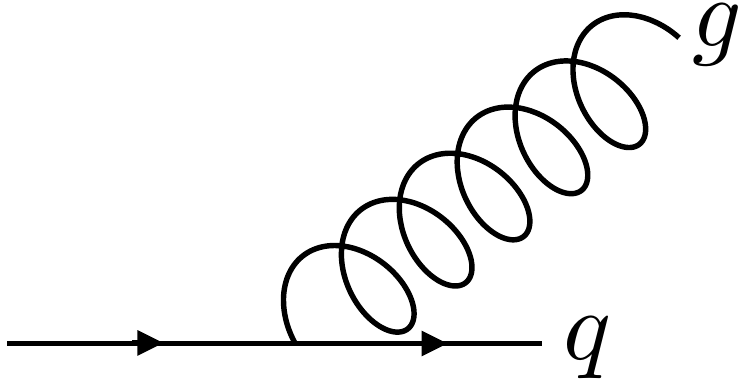}
\end{center}
or two soft and/or collinear gluons:

\begin{center}
\includegraphics[width=3.5cm]{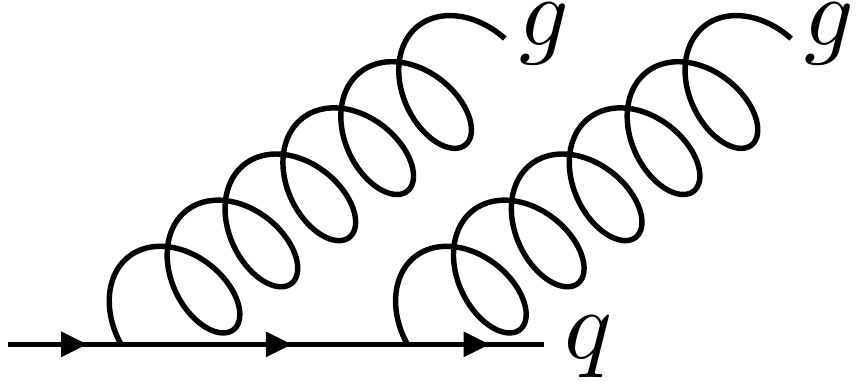}
\end{center}
or any number of soft and/or collinear gluons!  Every one of these configurations of soft and/or collinear gluons plus the quark have a divergent probability and all are degenerate with each other.  So how do we proceed?

Quantum mechanics saves us!  Feynman diagram perturbation theory is a degenerate perturbation theory.  And just like degenerate perturbation theory in quantum mechanics, we only get a finite result by summing over all degenerate configurations.  In quantum field theory, this result is called the KLN theorem \cite{Kinoshita:1962ur,Lee:1964is} (for Kinoshita, Lee and Nauenberg), extending results of Bloch and Nordsieck in quantum electrodynamics from the 1930s \cite{Bloch:1937pw}.  KLN ensures that summing over all individual divergent degenerate configurations produces a finite result.

We'll end this lecture by seeing how this is done.  Let's first go back to the probability distribution
\begin{equation}
P_{qg\leftarrow q}=\frac{2\alpha_s}{\pi}C_F\, \frac{dz}{z}\frac{d\theta}{\theta} = \frac{2\alpha_s}{\pi}C_F\, d\left(\log\frac{1}{z}\right)\,d\left(
\log\frac{1}{\theta}
\right)\,.
\end{equation}
On the right, I just re-expressed the probability as flat logarithmically in $z$ and $\theta$.  Thus, the phase space in the $(\log1/z,\log1/\theta)$ coordinates is a semi-infinite region where $\log1/z,\log1/\theta>0$.  Further, the probability distribution is flat: that is, there is uniform probability for the emission to be anywhere on the plane.  Thinking with degeneracy and KLN in mind, we can generalize this and say that the (arbitrary) gluon emissions off of a quark uniformly fill out the plane.  Representing each emission by a dot, for emissions off of a quark, we might have:

\begin{center}
\includegraphics[width=6cm]{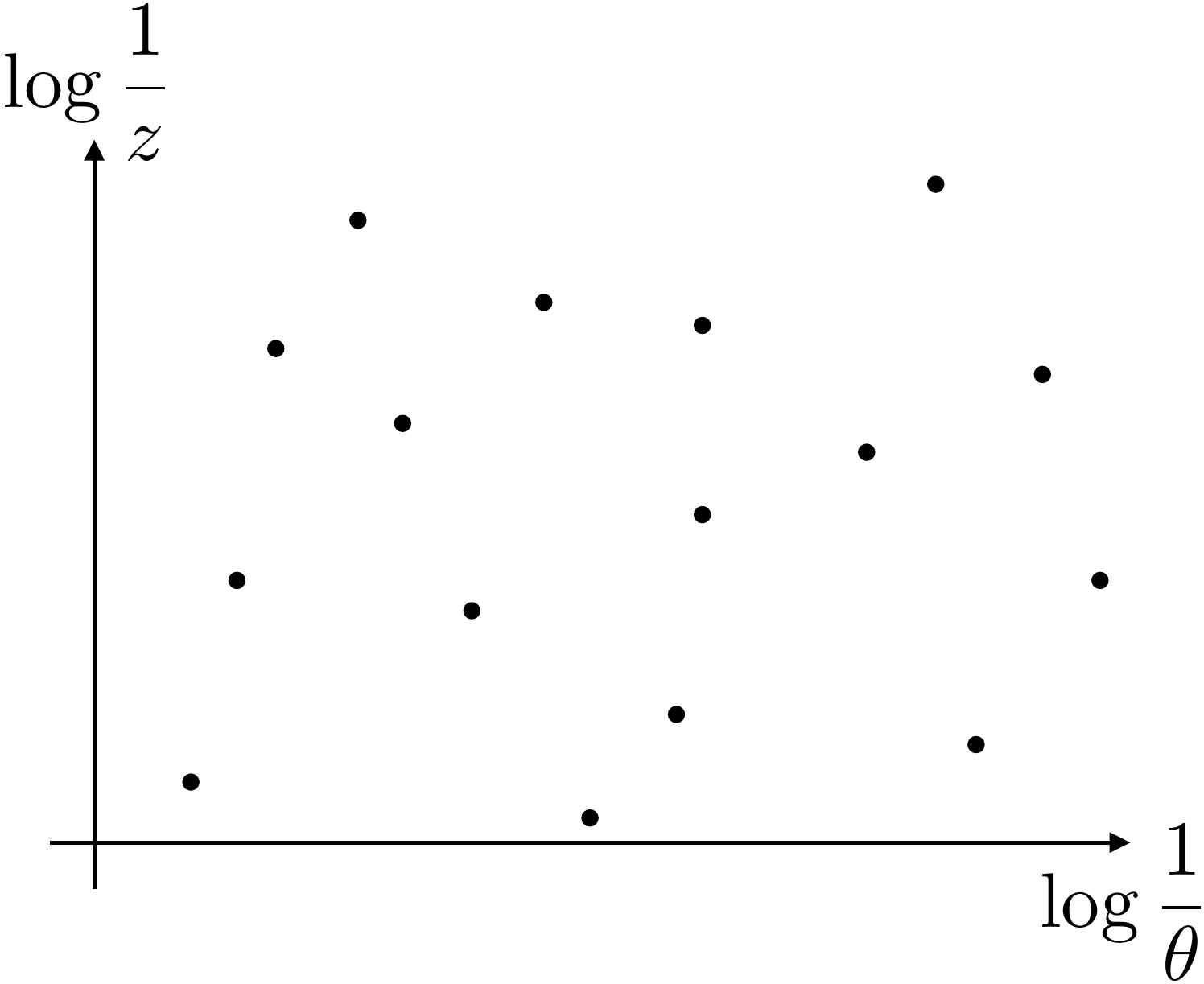}
\end{center}
Uniform in this logarithmic plane means exponentially far apart in ``real'' space, so gluon emissions off of a high energy quark are dominantly soft and/or collinear.  That is precisely what a jet is.

This plane is called the Lund plane, after researchers in Sweden who introduced it for studying jets \cite{Andersson:1988gp}.  Also, this uniform emission distribution is the starting point of modern Monte Carlo event generators and parton showers.  A modern Monte Carlo, like Pythia \cite{Sjostrand:2006za,Sjostrand:2014zea}, Herwig \cite{Bahr:2008pv,Bellm:2015jjp} and Sherpa \cite{Gleisberg:2008ta,Bothmann:2019yzt}, contains significant physics beyond this uniform assumption, such as: running $\alpha_s$ (emissions increase as $z,\theta$ decrease), fixed-order corrections (corrections to $1/z,1/\theta$ distributions), cut off at the scale of hadron masses, etc.  Nevertheless, while simple, this uniform emission phase space has a lot of physics.  In the next lecture, we will explore the effect of some of these subdominant corrections.

Before we do a calculation, I want to orient you in the Lund plane.  Let's draw it again:

\begin{center}
\includegraphics[width=7.75cm]{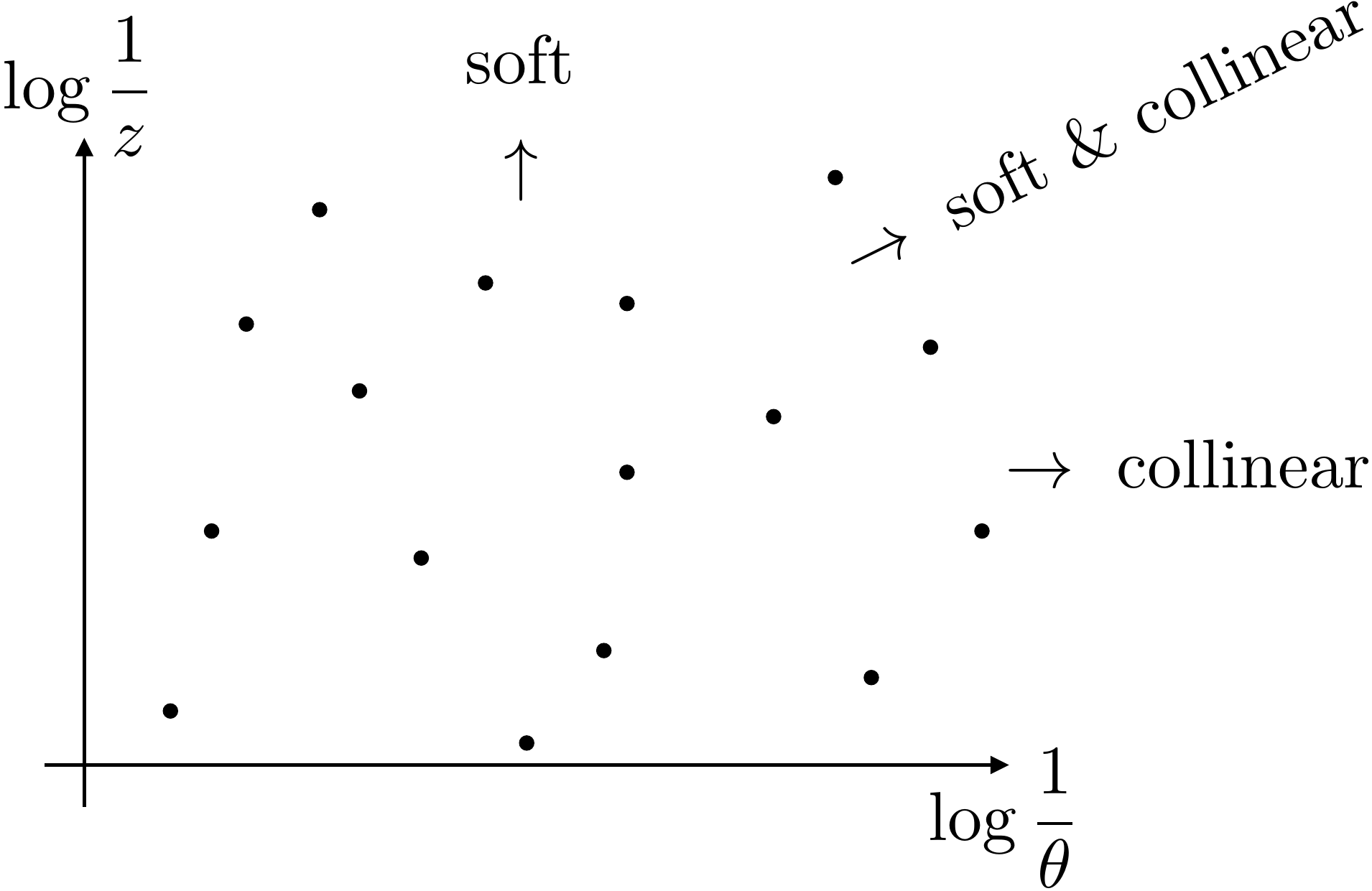}
\end{center}
I've also included regions to guide the eye.  The origin, where $z,\theta\sim 1$, corresponds to high-energy, wide-angle gluon emission.  The degenerate limits live off at $\infty$, with different physical origins for different infinities.  Vertical in the plane, $z\to 0$, is the soft limit, horizontal is the collinear limit $\theta \to 0$, and diagonal is a correlated soft and collinear limit.

With this picture in place, let's now calculate the distribution of a particular observable called an angularity $\tau_\alpha$ \cite{Berger:2003iw,Almeida:2008yp,Ellis:2010rwa,Larkoski:2014uqa}.  The angularity can be defined in the energy fraction $z$/angle coordinates as
\begin{equation}
\tau_\alpha = \sum_{i\in J} z_i\theta_i^\alpha\,,
\end{equation}
where the sum runs over all particles $i$ in the jet $J$ (collections of emissions), and $\alpha > 0$ is an exponent that weights contributions from different angles.  We require that $\alpha > 0$ to ensure that the observable is infrared and collinear safe: as we will see, this means that arbitrarily soft or collinear emissions cannot contribute to the observable (at least not dominantly so).  One important thing to note is that there is no preferred ordering to emissions of a scale-invariant system.  ``Scale invariant'' means that any scale we impose on the system can and will exist in that system.  We will see how measuring the angularities sets one particular ordering.

We had mentioned before that uniform logarithmically means exponentially far in real space.  Because each term in the definition of the angularities is weighted by a product of energy and angle, there will be one emission in the jet that dominates the value of $\tau_\alpha$, and all others will be exponentially small.  So, with one emission dominating, note that
\begin{equation}
\tau_\alpha = z\theta^\alpha\,,
\end{equation}
or
\begin{equation}
\log\frac{1}{\tau_\alpha} = \log\frac{1}{z}+\alpha \log\frac{1}{\theta}\,,
\end{equation}
and so a fixed value of $\tau_\alpha$ corresponds to a straight line on the $(\log1/z,\log1/\theta)$ plane.  We can draw this as:

\begin{center}
\includegraphics[width=7.5cm]{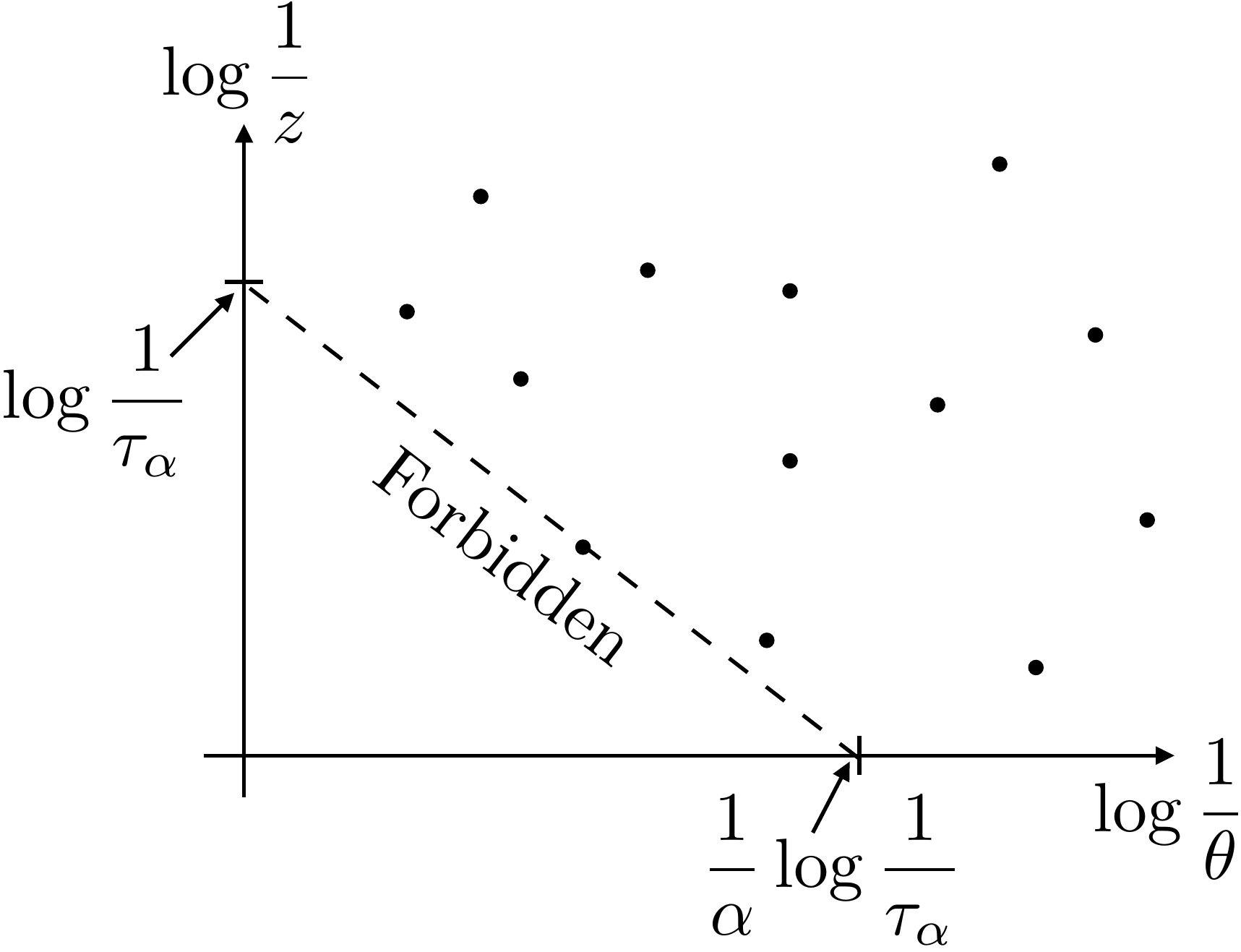}
\end{center}
We have drawn the abscissa and ordinate intercepts for the fixed value of $\tau_\alpha$ represented by the dotted line.  There is one emission on the dotted line, and all other emissions in the jet lie above the line.  Indeed, emissions below the dotted line are forbidden, as they would act to increase the value of $\tau_\alpha$ to be larger than what was measured.

So, for the measured value of $\tau_\alpha$, we must forbid all emissions at any point in the triangle below the dotted line.  To calculate the probability that the measured value of $\tau_\alpha$ is not larger than its value, we will do the following.  We must forbid emissions everywhere in the triangle, so let's isolate it and break it up into many pieces:

\begin{center}
\includegraphics[width=4cm]{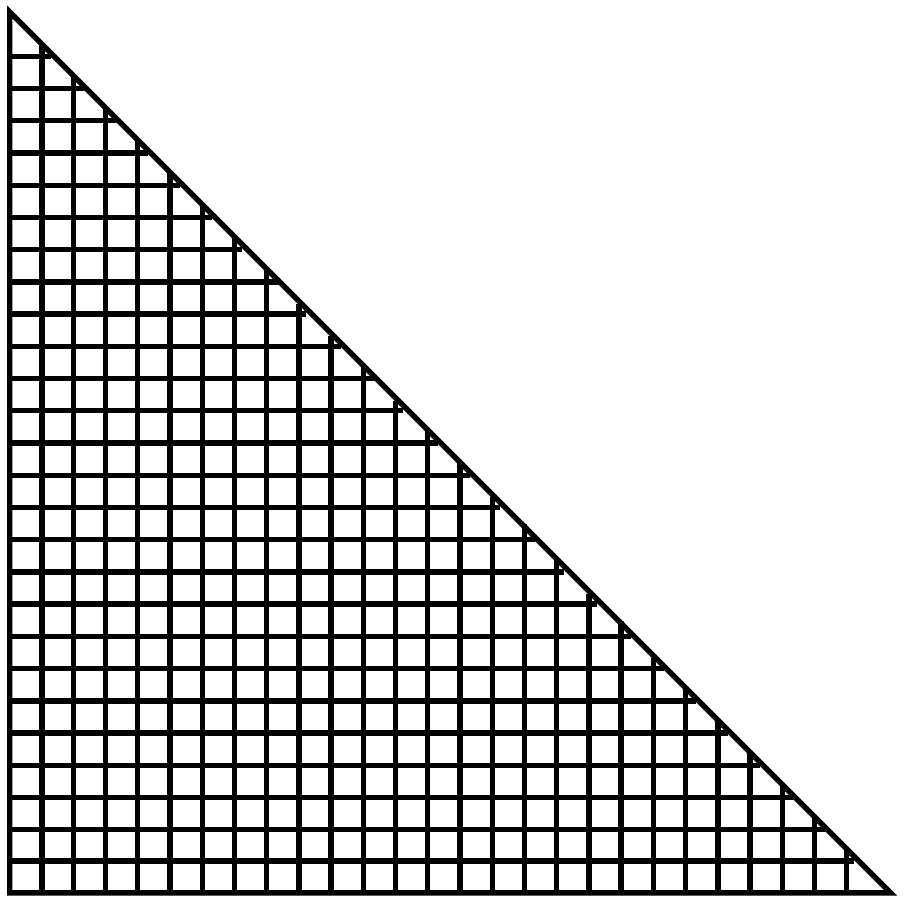}
\end{center}
There can be no emissions in any of the subregions.  This is an ``and'' statement in probability, so we must multiply the probability for no emissions in all regions together.  Let's break the triangle into $N$ equal area regions.  Then, the probability for an emission in one of the regions is uniform and equal to
\begin{equation}
\text{Probability for emission} = \frac{2\alpha_s}{\pi}C_F \frac{\includegraphics[width=0.5cm]{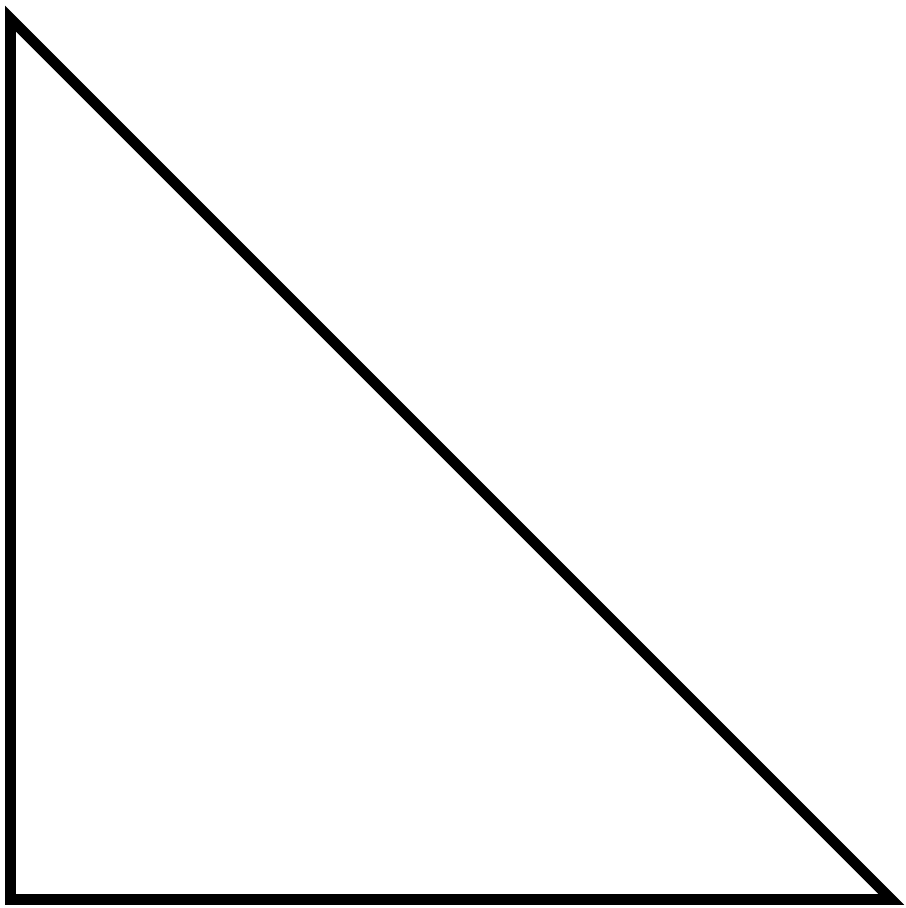}}{N}
\end{equation}
where \includegraphics[width=0.5cm]{triangle} is the area of the forbidden triangle:
\begin{equation}
\includegraphics[width=0.5cm]{triangle} = \frac{1}{2\alpha}\log^2\tau_\alpha\,.
\end{equation}
The, the probability for no emission in one small region is one minus this:
\begin{equation}
\text{Probability for no emission} = 1-\frac{2\alpha_s}{\pi}C_F \frac{\includegraphics[width=0.5cm]{triangle}}{N} = 1-\frac{\alpha_s}{\pi}\frac{C_F}{\alpha}\frac{\log^2\tau_\alpha}{N}\,.
\end{equation}
Then, the probability for no emission anywhere in the triangle is the product of probabilities of no emissions anywhere in each subregion:
\begin{equation}
P(\tau_\alpha \text{ less than measured value}) = \lim_{N\to\infty}\left(
1-\frac{\alpha_s}{\pi}\frac{C_F}{\alpha}\frac{\log^2\tau_\alpha}{N}
\right)^N = \exp\left[
-\frac{\alpha_s}{\pi}\frac{C_F}{\alpha}\log^2\tau_\alpha
\right]\,.
\end{equation}
The product transmogrifies into an exponential!  This exponential factor is called the Sudakov form factor \cite{Sudakov:1954sw}, and is simply a manifestation of the scale-invariant Poisson process of particle emission in high-energy QCD.

From a probability perspective, this Sudakov form factor is the cumulative probability distribution or CDF of $\tau_\alpha$.  To find the (differential) probability distribution function, we just differentiate:
\begin{equation}
p(\tau_\alpha) = \frac{d}{d\tau_\alpha}\exp\left[
-\frac{\alpha_s}{\pi}\frac{C_F}{\alpha}\log^2\tau_\alpha
\right] = -\frac{2\alpha_s}{\pi}\frac{C_F}{\alpha}\frac{\log\tau_\alpha}{\tau_\alpha}\exp\left[
-\frac{\alpha_s}{\pi}\frac{C_F}{\alpha}\log^2\tau_\alpha
\right]\,.
\end{equation}
This probability is normalized on $\tau_\alpha\in[0,1]$.  The issues with divergences with any fixed number of gluon emissions has been transformed into exponential suppression with the Sudakov form factor.

That's it for today---we'll use this intuition to understand aspects of machine learning later.  Next lecture, we'll introduce more complications into our simple picture of a jet.  Below are a couple of exercises.

\subsection{Exercises}

\begin{enumerate}

\item Consider the measurement of two angularities, $\tau_\alpha$ and $\tau_\beta$, with, say $\alpha> \beta$.  Calculate the Sudakov form factor for two angularities, the joint probability distribution $p(\tau_\alpha,\tau_\beta)$.  Further, ensure that the joint probability distribution marginalizes to the correct single probability distributions.  That is,
\begin{equation}
\int_{\tau_0}^{\tau_1}d\tau_\beta\, p(\tau_\alpha,\tau_\beta) = p(\tau_\alpha)\,,
\end{equation}
for particular bounds $\tau_0< \tau_\beta < \tau_1$.  For a hint to this problem, see Ref.~\cite{Larkoski:2013paa}.

\item (This is an extension of Exercise 9.3 in Ref.~\cite{Larkoski:2019jnv}.)  The ALEPH experiment at the Large Electron-Positron Collider (LEP) measured the number of jets produced in $e^+e^-\to$ hadrons collision events.  The experiment counted $n$ jets, if, for every pair $i,j$ of jets the following inequality is satisfied:
\begin{equation}
2\min[E_i^2,E_j^2](1-\cos\theta_{ij}) > y_{\text{cut}} E_{\text{cm}}^2\,,
\end{equation}
for $y_\text{cut}<1$, $E_i$ is the energy of jet $i$, $\theta_{ij}$ is the angle between jets $i$ and $j$ and $E_\text{cm}$ is the center-of-mass collision energy.  In the soft and collinear limits, determine the probability $p_n$ for observing $n$ jets, as a function of $y_\text{cut}$.

Note that the minimum number of jets is 2 ($e^+e^- \to q\bar q$) and gluons can be emitted from either the quark or the anti-quark.  Compare your result to figure 7 of Ref.~\cite{Heister:2003aj}.  What value of $\alpha_s$ fits the data the best?  This fit is imperfect because we're omitting a lot of important physics, but it will be qualitatively close.

\end{enumerate}

\clearpage

\section{Lecture 3: Jets at Higher Orders}

Welcome back to the third lecture on jet physics!  In this lecture, we will extend the results from the second lecture to include more physics to make our description of jets more realistic.  The first thing I want to address, though, is the property of infrared and collinear safety of the angularities.  We had found that the Sudakov form factor for the angularity measured on a quark jet took the form:
\begin{equation}
\Sigma_q(\tau_\alpha) = \exp\left[
-\frac{\alpha_s}{\pi}\frac{C_F}{\alpha}\log^2\tau_\alpha
\right]\,,
\end{equation}
for $\tau\in[0,1]$ and $\alpha > 0$.  Recall that $\alpha$ was the angular exponent of the angularity:
\begin{equation}
\tau_\alpha = \sum_{i\in J}z_i \theta_i^\alpha\,,
\end{equation}
and I said that it must be positive for infrared and collinear safety.  Using our axiom of scale invariance of QCD, we were lead to a probability for single gluon emission that diverged in the soft (= low energy) and collinear limits.  When all degenerate configurations of emitted gluons are summed up, we generate the Sudakov form factor, which is finite by the KLN theorem.  However, the Sudakov form factor is a series with terms to all orders in the coupling $\alpha_s$.  What if we just want a description of QCD jets to a fixed order in $\alpha_s$?  That is, a description as provided by some collection of Feynman diagrams?  Well, for the result to be sensible (i.e., finite) even though the fundamental probability distribution diverges in the soft and/or collinear limits, requires a delicate property of the observable that we choose to measure on the jet.  In particular, for the distribution of an observable to be finite almost everywhere in its domain requires that the soft and collinear limits of that observable map to a unique value.  That is, there is a single value of the observable for which all divergences from soft or collinear gluon emission are located.  Another way to state this criteria is that exactly 0 energy or exactly collinear gluons do not affect the value of the observable \cite{Ellis:1991qj}.  Such an observable for which this is true is called ``infrared and collinear safe'' or IRC safe.  Isolating all divergences to a single value of the observable means that away from that value, everything is well-defined and finite.

The angularities are IRC safe because soft and collinear gluons do not contribute to $\tau_\alpha$, for $\alpha > 0$.  However, not all possible observables or questions you can ask of a jet are IRC safe.  Perhaps the canonical example of non-IRC safety is that of multiplicity, or the number of particles in a jet.  A jet could consist of a single, bare quark so multiplicity would be 1.  However, say that a quark emits an exactly collinear gluon; now multiplicity would be 2.  However, this exactly collinear emission is degenerate to the bare quark, and violates the assumptions of KLN for finiteness.  So, multiplicity is not IRC safe.

Another thing to address is what the Sudakov form factor is for a high-energy gluon jet.  From what we have discussed thus far, the only difference between a quark and gluon is the color that either carry.  (Implicitly also their spin is importantly different, and we will address one consequence of that shortly.)  A quark is in the fundamental representation of SU(3) color, and so the number of colors it can share with a soft or collinear gluon is controlled by the fundamental quadratic Casimir, $C_F = 4/3$.  Gluons, by contrast, live in the adjoint representation of SU(3) color, and can share more color with soft or collinear gluons than can quarks.  The adjoint quadratic Casimir $C_A = 3$ in QCD, so this has consequences for the gluon Sudakov factor and a heuristic for understanding properties of QCD jets.  The gluon Sudakov is found by replacing $C_F \to C_A$ in the quark Sudakov:
\begin{equation}
\Sigma_g(\tau_\alpha) = \exp\left[
-\frac{\alpha_s}{\pi}\frac{C_A}{\alpha}\log^2\tau_\alpha
\right]\,.
\end{equation}
Because $C_A > C_F$, gluons are more likely to emit soft/collinear gluons, so it is more difficult for a gluon jet to have a very small value of $\tau_\alpha$.  Hence, the Sudakov provides a greater exponential suppression for gluons than quarks.

This distinction between the factor in the exponent of the Sudakov form factor for quark and gluon jets can also be understood from perhaps a more familiar phenomenon: radioactive decay.  The color factors $C_F$ and $C_A$ control the rate of particle emission from a quark or gluon, respectively, in the same way that the decay rate of an unstable particle (the inverse of its half-life) controls the rate of decay.  The fact that $C_A > C_F$ means that there is greater probability for a gluon to emit more particles than a quark for the same value of the angularity:
\begin{center}
\includegraphics[width=5cm]{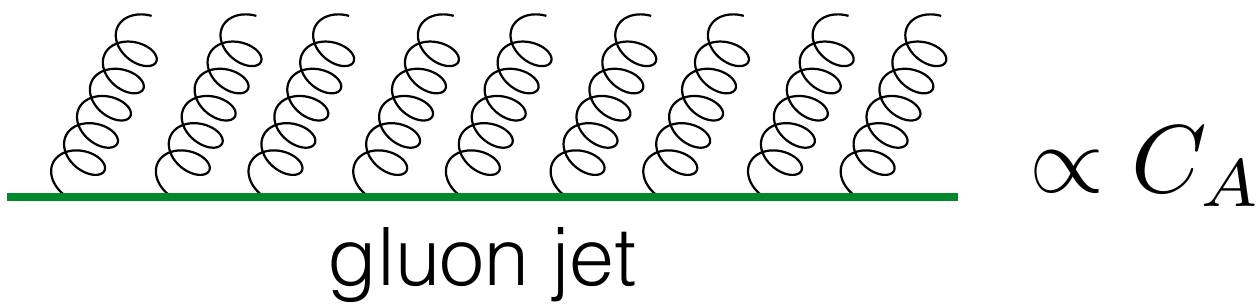} \hspace{3cm}
\includegraphics[width=6cm]{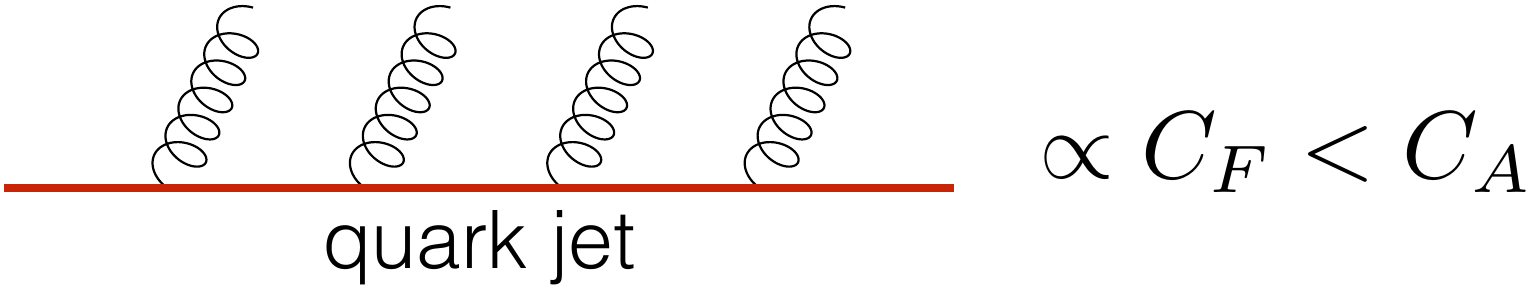}
\end{center}
Here, I have illustrated the relative size of the Casimirs $C_A$ and $C_F$ through the number of particles emitted off of the quark or gluon.  Note that $C_A/C_F = 9/4$ in QCD, and so, roughly, for every 9 particles a gluon emits, a quark only emits 4.  With more emissions in a gluon jet than a quark jet, it is exponentially less likely that the value of the angularity on a gluon jet is less than that on a quark jet.  This will be an important point next lecture.

For the remainder of this lecture, I want to survey a few physics effects that modify the simple, scale-invariant picture we had developed of emissions in a jet.  I hinted at some of these things in the previous lecture, but we can be more quantitative.  The first, most important, phenomenon that needs to be included is the running of the strong coupling, $\alpha_s$.  The strength with which quarks and gluons interact with one another depends on the energy scale at which that interaction is probed.  Quantum mechanical effects are responsible for modifying the measured value of the strong coupling at a given energy scale through a mechanism called renormalization.  Calculating the renormalized coupling from first principles in QCD is beyond these lectures, but a physical picture is actually quite simple.  Renormalization is just a technique for controlled, quantitative ignorance, and can be exploited in numerous contexts.  A simple, classical mechanics, example of renormalization is of a mass on a spring immersed in a viscous fluid.  To determine the specific dynamics of the mass as it oscillates in the fluid can be extremely challenging, and may require equations of hydrodynamics.  However, we can instead develop an effective picture of the mass in the fluid by introducing an effective mass that, say, depends on the velocity with which the mass moves.  This velocity-dependent mass absorbs all of the complications of hydrodynamics into a few parameters that can be subsequently measured.  The running, or energy scale dependence, of $\alpha_s$ is exactly similar: we don't need to know the intricacies of QCD excitations of any wavelength to determine an effective description of the coupling.

This running of $\alpha_s$ is quantified by the so-called $\beta$-function, where we define
\begin{align}\label{eq:betadiffeq}
\mu^2\frac{\partial \alpha_s}{\partial \mu^2} = \beta(\alpha_s)\,.
\end{align}
Here, $\mu$ is an energy scale appropriate for the problem at hand.  The $\beta$-function can be calculated as a Taylor series in $\alpha_s$, and the first term is \cite{Gross:1973id,Politzer:1973fx}
\begin{align}
\beta(\alpha_s) = - \frac{\alpha_s^2}{4\pi}\left(
\frac{11}{3}C_A -\frac{4}{3}T_R  n_f
\right)+{\cal O}(\alpha_s^3)\,.
\end{align}
Here, $T_R = 1/2$ and quantifies a normalization choice for the SU(3) color matrices of QCD.  $n_f$ is the number of active quarks; that is, the number of quarks whose masses are less than the energy scale $\mu$.  With the value of the coupling at a reference scale $\mu_0$, the differential equation of Eq.~\ref{eq:betadiffeq} can be solved and the running coupling is
\begin{align}\label{eq:runningassol}
\alpha_s(\mu^2) = \frac{\alpha_s(\mu_0^2)}{1+\frac{\alpha_s(\mu_0^2)}{4\pi}\left(
\frac{11}{3}C_A -\frac{4}{3}T_R  n_f
\right)\log\frac{\mu^2}{\mu_0^2}}\,.
\end{align}
Note that the $\beta$-function is negative, which means that as the energy $\mu$ increases, the value of the coupling decreases.  In the asymptotic limit where $\mu\to \infty$, the coupling vanishes, and quarks and gluons become free, non-interacting particles.  This property of QCD is therefore referred to as ``asymptotic freedom''.

The most precise scale at which $\alpha_s$ has been measured is in $e^+e^-$ collisions at a center-of-mass of the $Z$ boson mass, $m_Z = 91.2$ GeV, where \cite{ParticleDataGroup:2020ssz}
\begin{align}
\alpha_s((91.2\text{ GeV})^2) = 0.118\,.
\end{align}
To get a sense for the rate of the change in the value of the coupling, the energy scales at which the value of the coupling increases or decreases by a factor of 2 from its value at the mass of the $Z$ boson are:
\begin{align}
\alpha_s(\mu^2) = \frac{\alpha_s((91.2\text{ GeV})^2)}{2} \qquad &\rightarrow \qquad \mu = 95\text{ TeV}\,,\\
\alpha_s(\mu^2) = 2\alpha_s((91.2\text{ GeV})^2) \qquad &\rightarrow \qquad \mu = 2.8\text{ GeV}\,.\nonumber
\end{align}
That is, at the very high energies probed by the LHC, the rate of change of the coupling is quite slow.  However, if your observable is sensitive to very low energy emissions, the value of the coupling increases rapidly.

The running coupling therefore induces a non-uniform structure of emissions on the Lund plane.  The only Lorentz-invariant quantity that the coupling can be sensitive to is the invariant mass $m^2$ of two particles after a splitting, and so the value of $\alpha_s$ is evaluated at this scale:
\begin{center}
\includegraphics[width=3cm]{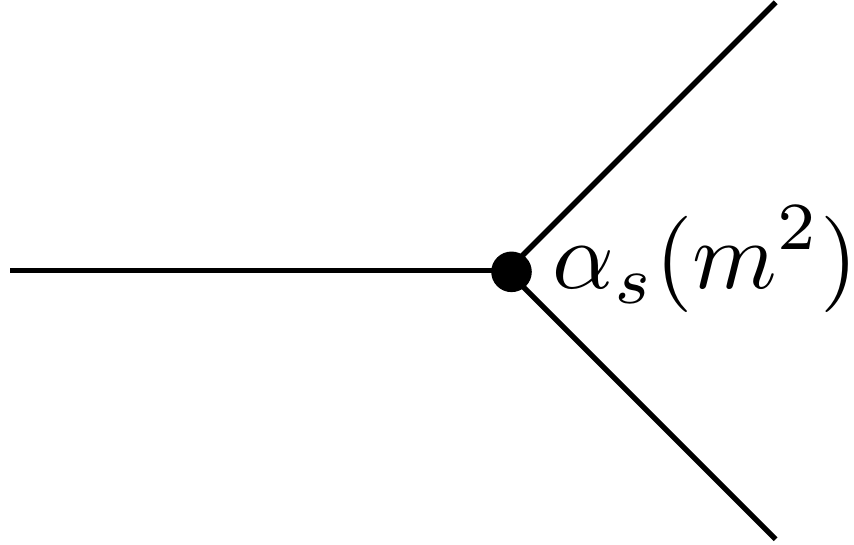}
\end{center}
As one is sensitive to lower and lower mass emissions on the Lund plane, $\alpha_s$ increases, and so the rate of particle emission correspondingly increases.  We can illustrate this by increasing the density of emissions on our image of the Lund plane, as we move to lower masses (up and to the right):
\begin{center}
\includegraphics[width=7cm]{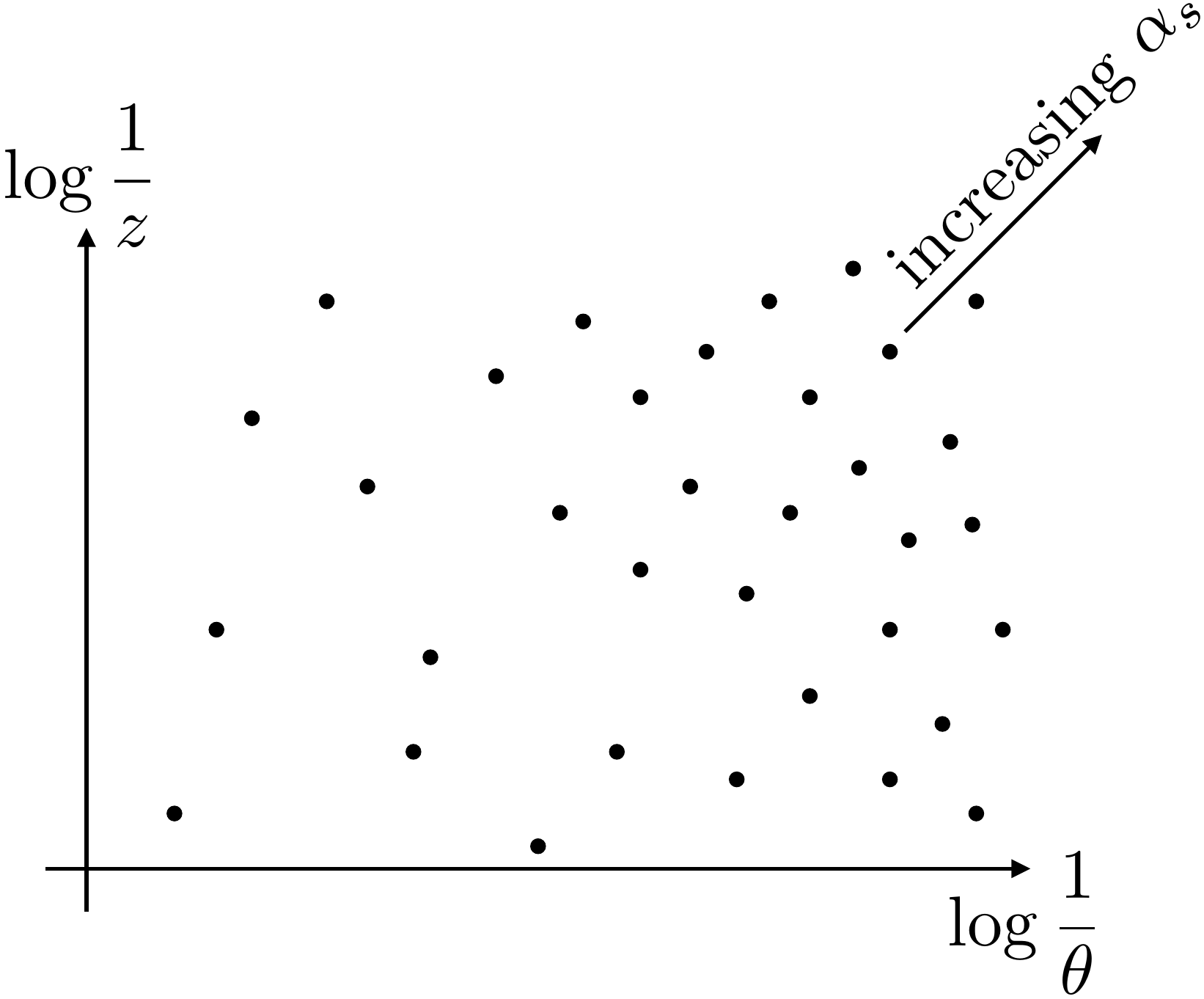}
\end{center}

Next, in our simple construction of the probability of gluon emission from a quark, for example, we assumed an invariance to rescaling of the gluon energy fraction.  This was motivated by the 0 energy limit, in which the energy of a 0 energy gluon could be rescaled by an arbitrary factor $\lambda > 0$ and nothing changes.  However, this is clearly a limited approximation, because a gluon with non-zero energy will, at least, carry away energy from the quark, conserving the total quark-gluon energy.  By contrast, in our earlier approximation, we also assumed that the distribution of the angle between the gluon and quark exhibited a scaling symmetry, as well.  However, unlike the energy fraction, we can retain this collinear approximation, simply by requiring that the particles are close to one another in angle.  Within QCD, we can calculate the probability for collinear gluon emission from a quark and the resulting distributions are referred to as splitting functions.  The full energy dependence of the splitting functions further affects the density of emissions on the Lund plane, with collinear emissions modified from uniform logarithmically in energy.

The splitting functions from both an initial quark and an initial gluon have been calculated \cite{Altarelli:1977zs}.  For a quark splitting to a quark and a gluon, we have
\begin{align}
P_{qg\leftarrow q} = \frac{\alpha_s C_F}{\pi}\frac{1+(1-z)^2}{z}\, dz \,\frac{d\theta}{\theta}\,,
\end{align}
where $z$ is the energy fraction of the gluon, $z\in[0,1]$, and $\theta$ is the angle between the final quark and gluon.  In the limit in which the gluon is much lower energy than the quark, $z\to 0$, this reduces to
\begin{align}
P_{qg\leftarrow q} \to \frac{2\alpha_s C_F}{\pi}\frac{dz}{z} \,\frac{d\theta}{\theta}\,,
\end{align}
exactly as we had constructed from scale invariance in Eq.~\ref{eq:scalesplit}.  For an initial gluon, the complete splitting function has a couple different parts.  In the soft limit, we had identified gluon emission off of a high-energy gluon, and this will need to be generalized.  However, the gluon can also split to a quark--anti-quark pair, in an exactly analogous way that a photon can split into an electron and positron.  There is no enhanced soft limit for this splitting because the case when a quark becomes soft is not a degenerate limit, however, which is why we did not see it earlier.  When accounted for, the complete gluon splitting function is
\begin{align}
P_{ij\leftarrow g} = \frac{\alpha_s}{\pi} \left[
2C_A\left(
\frac{1-z}{z}+\frac{z}{1-z}+z(1-z)
\right)+T_R n_f \left(
z^2 + (1-z)^2
\right)
\right]dz\, \frac{d\theta}{\theta}\,.
\end{align}
It is interesting to note that this splitting function is symmetric for $z\leftrightarrow 1-z$, corresponding to an interchange of the final state particles.  For two final state gluons, this is required by Bose symmetry, and for a final state quark--anti-quark pair, this follows because a quark and anti-quark interact with a gluon in essentially the same way.

\begin{figure}
\begin{center}
\includegraphics[width=0.45\textwidth]{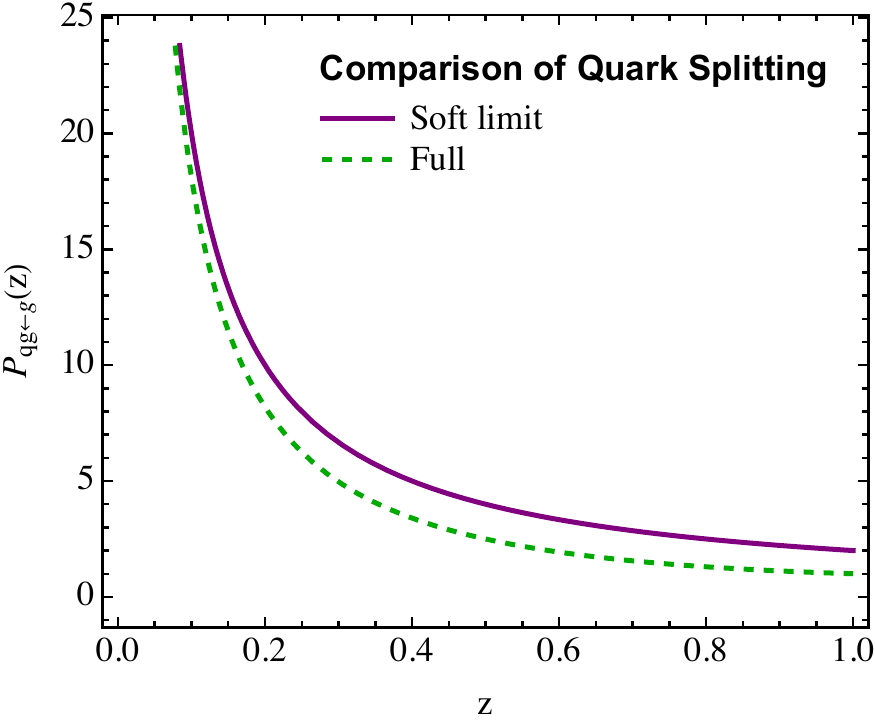}\ \ \ 
\includegraphics[width=0.45\textwidth]{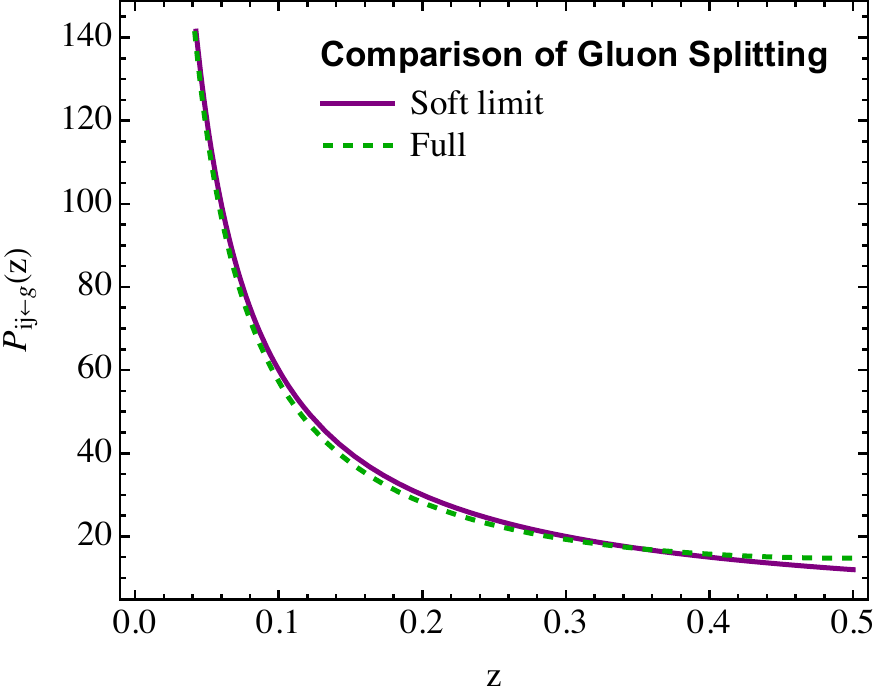}
\caption{\label{fig:splitfunccomp}
A comparison of the energy fraction distribution in a $1\to 2$ splitting between the soft limit and full splitting function.  The quark (gluon) splitting is on the left (right) and the number of active quarks has been set to $n_f = 5$.}
\end{center}
\end{figure}

To observe the effect of including terms in the splitting functions beyond the soft limit, we have plotted the energy fraction $z$ dependence of the quark and gluon splitting functions in Fig.~\ref{fig:splitfunccomp}.  Here, the soft limit result is plotted in solid purple, and is just proportional to $1/z$ (which is why it diverges as $z\to 0$).  The full splitting functions, quoted above, are plotted as dashed green in the plot.  For quarks, the full splitting functions produce a significantly softer distribution than just $1/z$; it is less likely for the gluon to take away most of the initial quark's energy from the full splitting function as compared to the soft limit.  Interestingly, the full splitting function for the gluon is very close to just its soft limit, and this is mostly a numerical coincidence between the subleading terms for gluon or quark--anti-quark final states.  Note also that the soft limit $1/z$ does not have the correct physics as $z\to 1$.  In this limit, the other gluon becomes soft, but, as identical bosons, we cannot distinguish these limits.

The final point for this lecture is that this entire picture of the quasi-particles of a jet being quarks and gluons breaks down at sufficiently low energies.  There are a couple of different reasons why, but they have correlated physical origins.  First, our analysis of particle production has been strictly perturbative, in that we imagine an initial state and then perturb about it, emitting individual particles sequentially.  This approximation works well when the coupling $\alpha_s$ is small.  But, with our understanding of asymptotic freedom and its running, at low enough energies we can no longer say that $\alpha_s$ is small.  Given the expression for the running coupling in Eq.~\ref{eq:runningassol} and its value at $\mu_0 =m_z= 91.2$ GeV, we can determine the energy $\Lambda$ at which the coupling would (ostensibly) diverge.  This energy scale is called the Landau pole \cite{Landau:1954cxv} and has value
\begin{align}
\Lambda = m_Z \exp\left[
-\frac{2\pi}{\alpha_s(m_Z^2)}\frac{1}{
\frac{11}{3}C_A -\frac{4}{3}T_R  n_f}
\right] \simeq 246\text{ MeV}\,.
\end{align}
Here, we have set $n_f = 3$ because, with the benefit of foresight, there are three quarks with mass less than this scale.  So for our analysis to make any sense, the relevant energy scales in our jet must far exceed this value.

Additionally, this Landau pole occurs in an auspicious place, because the bound states of QCD, the hadrons, all have masses in the hundreds of MeV range.  For massless quarks and gluons to be good quasiparticles of our jets, the fact that the particles we actually detect are hadrons must be a subdominant effect.  The description of a jet must include hadrons when the invariant mass of a splitting becomes comparable to the masses of the hadrons:
\begin{center}
\includegraphics[width=3.5cm]{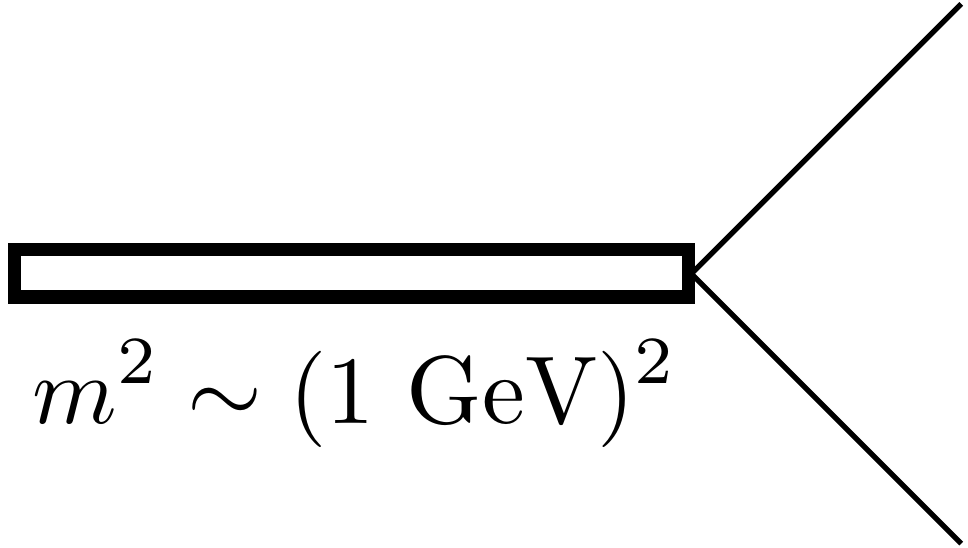}
\end{center}
If this is the case, then we cannot imagine that a quark split to a quark and a gluon, for example.  This diagram instead would describe something like the two-body decay of a hadron to other hadrons or leptons, and the masses of all particles involved would be important.  If we restrict attention to emissions that are sufficiently high energy to avoid the Landau pole and hadron masses, this effectively fixes a maximum distance from the origin that we can consider on the Lund plane.  If we get too far away, this simple, scale-invariant picture is completely wrong, and new quasi-particles, the hadrons, take over.

\subsection{Exercises}

\begin{enumerate}

\item With the complete splitting functions, we can get more interesting answers to questions we ask of jets.  For example, we can use the complete splitting functions to modify the Sudakov form factor and correspondingly the resummed distribution of the angularities.  Because we are working away from the soft limit, we must modify the definition of the angularity to ensure its IRC safety.  The generalized definition you can use here is
\begin{align}
\tau_\alpha = z(1-z)\theta^\alpha\,.
\end{align}
The Sudakov form factor for an angularity $\tau_\alpha$ takes the general form
\begin{align}
\Sigma(\tau_\alpha) = e^{-R(\log \tau_\alpha)}\,,
\end{align}
where $R(\log \tau_\alpha)$ is sometimes called the radiator function.  It is defined from the splitting functions as
\begin{align}
R(\log\tau_\alpha)\equiv \int_0^1 dz \int_0^1 \frac{d\theta}{\theta}\, P(z)\, \Theta(z(1-z)\theta^\alpha - \tau_\alpha)\,.
\end{align}
Here, we have just set the upper limit of the angular integral to 1 for concreteness, and $P(z)$ is the energy fraction dependence of the splitting functions.  For example, for quark splitting, this would be
\begin{align}
P_{qg\leftarrow q}(z) = \frac{\alpha_s C_F}{\pi}\frac{1+(1-z)^2}{z}\,.
\end{align}
Calculate the radiator, the Sudakov form factor, and the differential probability distribution for $\tau_\alpha$, on both quark and gluon jets.  In the radiator, only keep those terms that are logarithmic in the angularity $\tau_\alpha$.  For simplicity, assume that $\alpha_s$ is fixed, and does not run.  How do the full splitting functions modify the distribution we had calculated in the soft limit in the previous lecture?

\item Even with the soft limit as an approximation, it is illuminating to observe the modification of the Sudakov form factor with a running coupling, $\alpha_s(\mu^2)$.  The radiator in the soft limit but with a running coupling would be
\begin{align}
R(\log\tau_\alpha)\equiv 2C_i\int_0^1 \frac{dz}{z} \int_0^1 \frac{d\theta}{\theta}\, \alpha_s(\mu^2)\, \Theta(z\theta^\alpha - \tau_\alpha)\,.
\end{align}
In general, the scale of the coupling $\mu^2$ depends on the momenta of the final state particles.  Here, we will take this scale to be the invariant mass of the jet in the soft limit, where
\begin{align}
\mu^2 = E^2z\theta^2\,,
\end{align}
and $E$ is the fixed energy of the jet.  With the expression for the running coupling in Eq.~\ref{eq:runningassol}, calculate the radiator, Sudakov form factor, and the differential probability distribution for the angularity.  How does the running coupling modify the distribution we had calculated in the soft limit in the previous lecture?

\end{enumerate}

\clearpage

\section{Lecture 4: Machine Learning for Quark vs.~Gluon Discrimination}


In this final lecture, we're going to think like a machine to understand the problem of binary discrimination, or, techniques for distinguishing two samples mixed in an ensemble.  Machine learning is exploding as a discipline in particle physics \cite{Larkoski:2017jix,Guest:2018yhq,Radovic:2018dip,Albertsson:2018maf,Carleo:2019ptp,Bourilkov:2019yoi,Karagiorgi:2021ngt,Feickert:2021ajf}, so there is no hope for me to discuss broadly how it is employed.  However, my biased viewpoint is from that of a theorist who is selfishly interested in learning more about Nature.  So how can we use machine learning to learn more as flesh-and-blood humans?  Let's first define what we are working with.

My theorist definition of a machine (or neural network or any fancy computer science algorithm) is the following.  A machine is a black box which takes in input and returns output:

\begin{center}
\includegraphics[width=4.5cm]{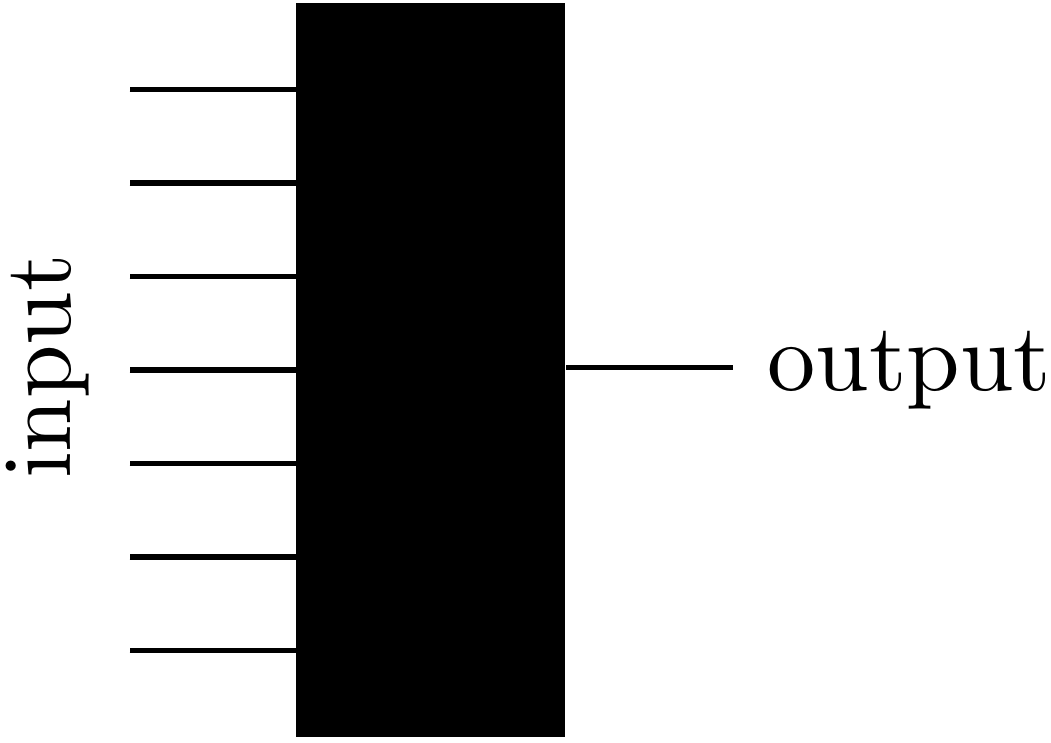}
\end{center}
By ``black box'', I mean that the machine performs some manipulation on the input to produce the output, but the way in which it does it is unknowable to me.  Don't worry; it will turn out fine that we don't know what the machine does.  I have also illustrated the input as multiple entries and the output as a single result.  That is, we consider the input to be some $n$ dimensional vector $\vec x$ and the output to be a single number $g$.  So all the machine is is a function of the input:
\begin{equation}
\text{Machine: }\mathbb{R}^n \to \mathbb{R}\,,
\end{equation}
which can be represented by the function $g\equiv g(\vec x)$.  One can consider more general inputs and outputs, but we will keep it restricted to this scalar function case, again for simplicity.

The first result that allows us to learn anything at all is the universal approximation theorem \cite{cybenko,hornik,leshno}.  For our purposes, the statement of the universal approximation theorem is the following.  A sufficiently large and powerful machine can output an arbitrary function of the inputs.  Concretely, let's say that the input data are entries of the vector $\vec x$, and the output of the machine is the function $g(\vec x)$:

\begin{center}
\includegraphics[width=4cm]{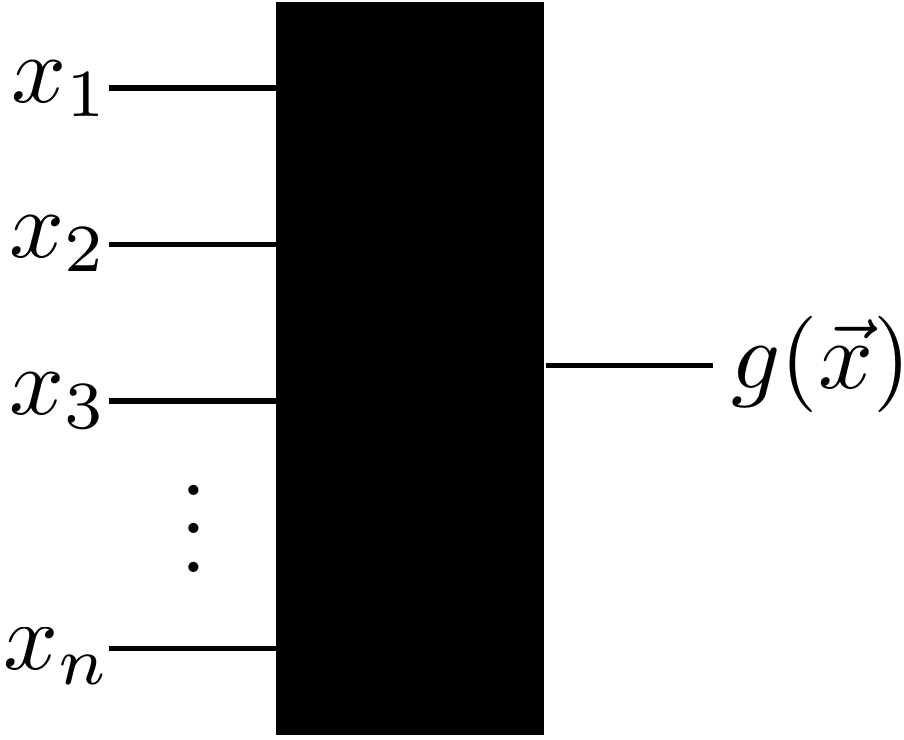}
\end{center}
The universal approximation theorem states that we can modify the inputs to (almost) any collection of functions $\{f_i(\vec x)\}_{i=1}^n$ and the machine will return (or ``learn'') the same function $g(\vec x)$:

\begin{center}
\includegraphics[width=4.5cm]{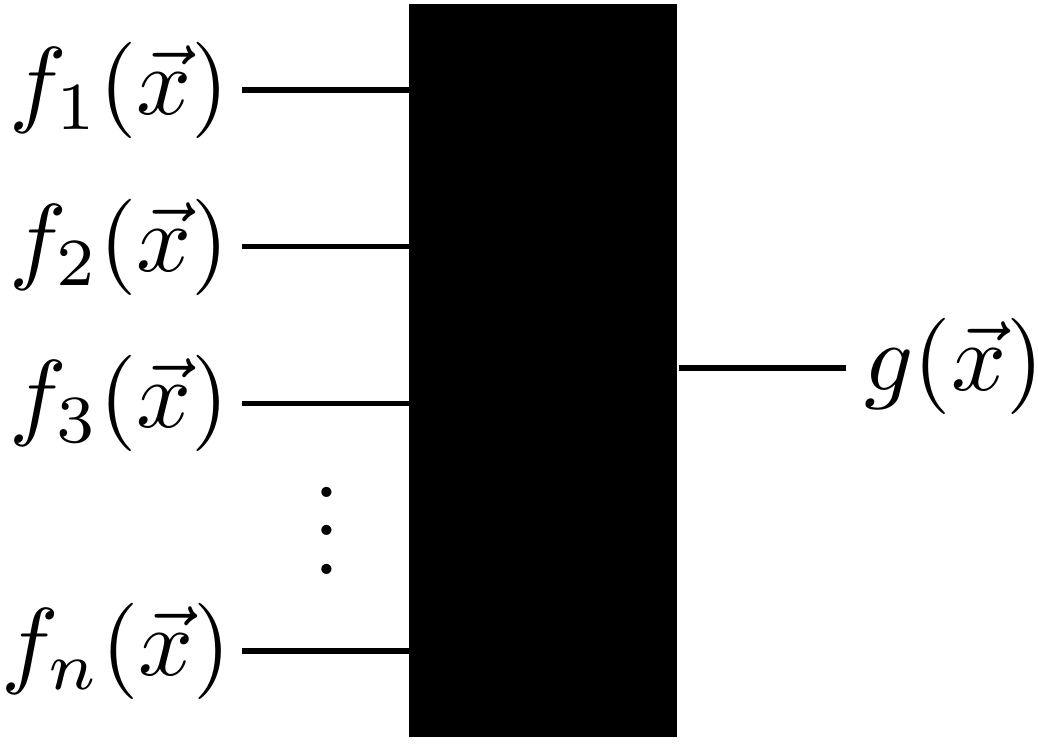}
\end{center}
As long as the collection of $f_i$ functions contains the same total information as the $\vec x$ input data, the machine will always learn $g(\vec x)$.  So, the universal approximation theorem states that we can choose a convenient form of input data, and so we can optimize for a form we understand well.

Now, a lot of machine learning practitioners will say that the universal approximation theorem is less than unuseful because the actual behavior of the machine depends so sensitively on precise implementation and architecture.  Further, in practice one only ever trains a machine on a finite amount of data, and so a sufficiently large and powerful machine can simply memorize the input data.  While this would result in perfectly analyzed training data, it would be extremely overtrained, and would fail to generalize to novel input data.  However, I again want to emphasize that we are not interested here in determining how a particular machine performs; we want to think like a machine to learn something about Nature.

The specific task we would like the machine to perform is binary discrimination, or, signal versus background discrimination.  The formulation of a binary discrimination problem is that we want the machine to separate, as efficiently as possible, signal from background events in a mixed ensemble.  What we mean by ``signal'' and ``background'' is that signal events, or a collection of the input data, are drawn from the probability distribution $p_s(\vec x)$, while background events are drawn from the probability distribution $p_b(\vec x)$.  Given a collection of identified signal and background events, the machine learns what the probability distributions are, and correspondingly outputs the probability with which it believes that an individual event is signal-type or background-type.

It doesn't do this separation blindly; we know the optimal binary discriminant from the result of the Neyman-Pearson lemma \cite{Neyman:1933wgr}.  In the 1930s, Jerzy Neyman and Egon Pearson proved that the likelihood ratio is the optimal binary discriminant.  The likelihood ratio ${\cal L}$ is simply the ratio of signal to background probability distributions:
\begin{equation}
{\cal L} = \frac{p_s(\vec x)}{p_b(\vec x)}\,.
\end{equation}
The likelihood naturally and beautifully separates signal and background, to the maximal extent provided by the probability distributions.  If ${\cal L}\to 0$, then the background probability is large compared to signal, and vice-versa for ${\cal L}\to \infty$.  Thus,
\begin{align}
&{\cal L}\to 0\qquad \Rightarrow \qquad \text{pure background}\,,\\
&{\cal L}\to \infty\qquad \Rightarrow \qquad \text{pure signal}\,.
\end{align}
Additionally, a larger class of quantities than strictly just the likelihood are equal in discrimination power.  Any function monotonic in the likelihood ${\cal L}$ is equivalent in discrimination power.  This monotonic function can be exploited to simplify the range of values that the discrimination observable assumes.  For example, the function
\begin{equation}
h({\cal L}) = \frac{{\cal L}}{1+{\cal L}} = \frac{p_s(\vec x)}{p_s(\vec x)+p_b(\vec x)}\,,
\end{equation}
is monotonic in ${\cal L}\in[0,\infty)$, but nicely maps to the domain $h({\cal L})\in[0,1]$.  In particular:
\begin{align}
&h({\cal L}\to 0)\to 0\qquad \Rightarrow \qquad \text{pure background}\,,\\
&h({\cal L}\to \infty)\to 1\qquad \Rightarrow \qquad \text{pure signal}\,.
\end{align}
Again, there's nothing special about this $h({\cal L})$, but it can be a nice object to consider.

So, for the case at hand, our machine takes in a collection of input data $\vec x$ for signal and background events drawn from probability distributions $p_s(\vec x)$ and $p_b(\vec x)$, respectively, and outputs the likelihood ratio, or a monotonic function of it:

\begin{center}
\includegraphics[width=5cm]{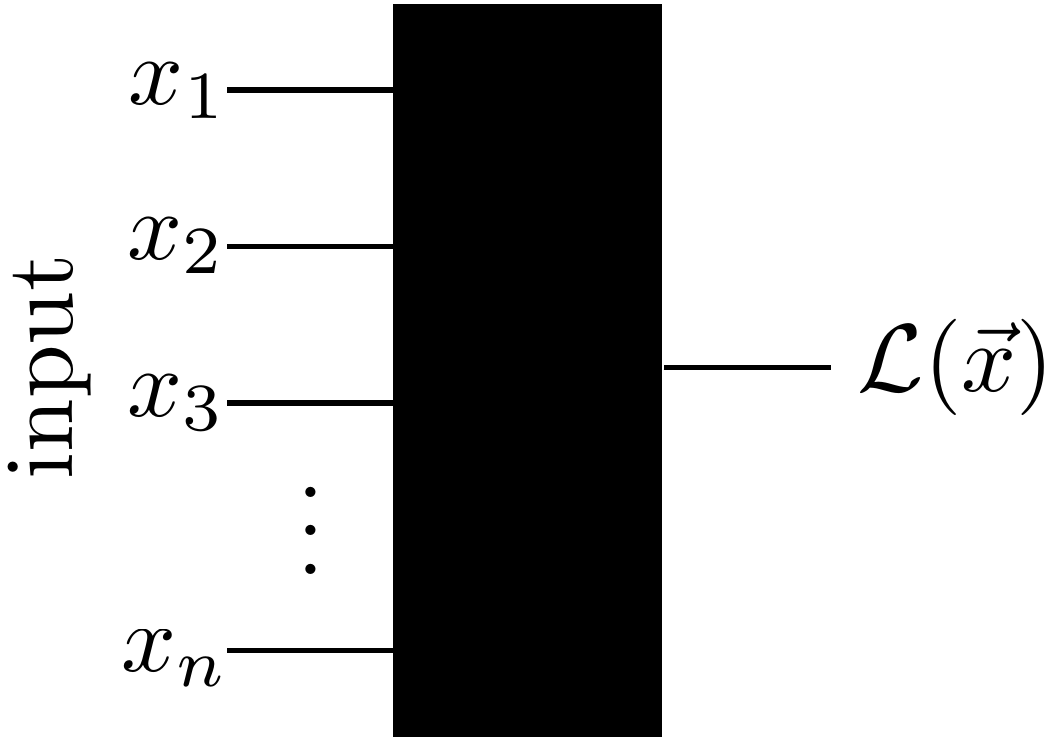}
\end{center}
With that set-up, let's now describe the physical system we would like to learn about.  Our goal in the rest of this lecture is to learn something about the likelihood ratio for the problem of discrimination of jets initiated by high-energy quarks versus those jets initiated by high-energy gluons.  The input data that we will use in our machine is the collection of four-momenta of particles in quark or gluon jets.  Let's call this set of momenta $\{p_i\}_{i\in J}$, and the quark and gluon probability distributions are thus:
\begin{equation}
p_q\left(\{p_i\}_{i\in J}\right)\,, \qquad
p_g\left(\{p_i\}_{i\in J}\right)\,,
\end{equation}
where $J$ is the jet of interest (i.e., the collection of soft and collinear particles).  The likelihood ratio is then
\begin{equation}
{\cal L} = \frac{p_g\left(\{p_i\}_{i\in J}\right)}{p_q\left(\{p_i\}_{i\in J}\right)}\,.
\end{equation}
So, there we go.  We're done, right?  While true, there is almost zero information in these statements; specifically, we know essentially nothing about the functional form of ${\cal L}$ for quark and gluon discrimination.  Can we learn any information that will help us in our task?

The first problem with this formulation is the explicit use of particle four-vectors as the input data.  This is a rather poor way to organize the information in a jet because, by the soft and collinear singularities of QCD, a substantial amount of particles will have (nearly) degenerate momenta, which is challenging to interpret theoretically so we humans can actually learn something.  So, using the universal approximation theorem, let's see if we can reorganize the information contained in four-vectors into a much more useful, and human-interpretable, form.

First, let's see what we are dealing with and what the dimension of the input space actually is.  A generic four-vector has four real components, so the four-vectors of $N$ particles (na\"ively) spans some real $4N$-dimensional space.  However, as real particles, their momenta are all on-shell.  For simplicity (but no other reason) let's assume that all particles are massless.  As such, an on-shell, massless four-vector actually only has 3 degrees of freedom, so the space of input momenta is only $3N$ real dimensions.  Additionally, total energy and momentum are conserved, which imposes 4 further linear constraints on all momenta.  Thus, there are $3N-4$ degrees of freedom to completely define the four-momenta of $N$ particles in a jet (assuming on-shellness and total momentum conservation).  So, we just need to identify $3N-4$ functions of the particles' momenta appropriately and we can use the universal approximation theorem to claim that our machine would find the same likelihood ratio.

So, what functions of momenta should we use?  This is a matter of taste, but if we want to exploit our perturbative understanding of QCD to the problem at hand, then we would want these functions to be infrared and collinear safe observables.  What $3N-4$ IRC safe observables should we use?  This is much more of a matter of taste, and there are many possible answers, but here we will just consider the $N$-subjettiness observables \cite{Thaler:2010tr,Thaler:2011gf}.  $N$-subjettiness is a class of IRC safe observables that extends the angularities to resolve $N$ prongs in a jet.  As a function of energy fractions $z_i$ and angles, $N$-subjettiness $\tau_N^{(\alpha)}$ is defined to be:
\begin{equation}
\tau_N^{(\alpha)} = \sum_{i\in J}z_i\, \min\left\{
\theta_{i1}^\alpha\,,\theta_{i2}^\alpha\,,\dotsc\,, \theta_{iN}^\alpha
\right\}\,,
\end{equation}
and $\alpha > 0$.  Here, $\theta_{iK}$ is the angle between particle $i$'s momentum and axis $K$ in the jet.  Defining $N$-subjettiness requires placing $N$ axes in the jet, nominally in the directions of dominant energy flow.  For example, consider $2$-subjettiness measured on a jet with two hard particles, 1 and 2, and one soft particle 3.  The two axes would, for example, align with particles 1 and 2 and only particle 3 would contribute to $\tau_2^{(\alpha)}$:
\begin{align}
\raisebox{-1.5cm}{\includegraphics[width=3cm]{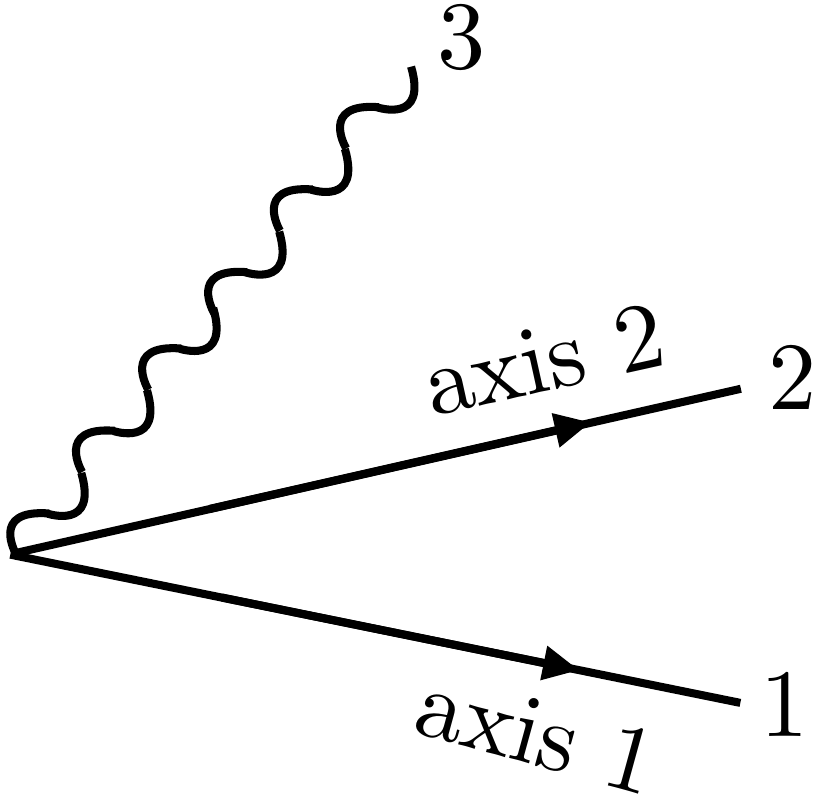}}\qquad  \Rightarrow \qquad \tau_2^{(\alpha)} &=\sum_{i\in J}z_i\, \min\{\theta_{i1}^\alpha,\theta_{i2}^\alpha\}=z_3 \min\{\theta_{31}^\alpha,\theta_{32}^\alpha\}\,.
\end{align}
Angularities are 1-subjettiness, and IRC safety of $N$-subjettinesses essentially follows from IRC safety of angularities (with some caveats regarding axes choice).  $N$-subjettinesses have the added benefit that they are additive, and so multiple soft and collinear emissions in the jet generates a Sudakov form factor, exactly as we observed for angularities.

So, we already know a lot about the space of $N$-subjettinesses, so we can choose $3N-4$ of them to resolve the four-momenta of $N$ particles in the jet.  Because time is short, I won't go into explicit details about what that collection of $N$-subjettinesses should be to ensure that they have the same information as the collection of four-vectors.  See Ref.~\cite{Datta:2017rhs} for all the details.

Now we're cooking.  What properties of the likelihood ratio ${\cal L}$ for quark versus gluon discrimination can we learn using the $N$-subjettiness variables as inputs to our machine (which in this case is our brains!)?  I'll present a simplified argument here, and more details can be found in Ref.~\cite{Larkoski:2019nwj}.  Let's just imagine, for simplicity, our entire input space was just that of $\tau_1^{(2)}$ and $\tau_2^{(2)}$, the 1- and 2-subjettinesses with angular exponents equal to 2:
\begin{align}
&\tau_1^{(2)} =\sum_{i\in J}z_i \theta_i^2\,, &\tau_2^{(2)} = \sum_{i\in J}z_i\, \min\{\theta_{i1}^2,\theta_{i2}^2\}\,.
\end{align}
Note that both $\tau_1^{(2)},\tau_2^{(2)}>0$ and $\tau_1^{(2)}>\tau_2^{(2)}$ because with two axes in the jet the distance to any particle to those axes is less than or equal to the distance to a single axis in the center of the jet:

\begin{center}
\includegraphics[width=8.5cm]{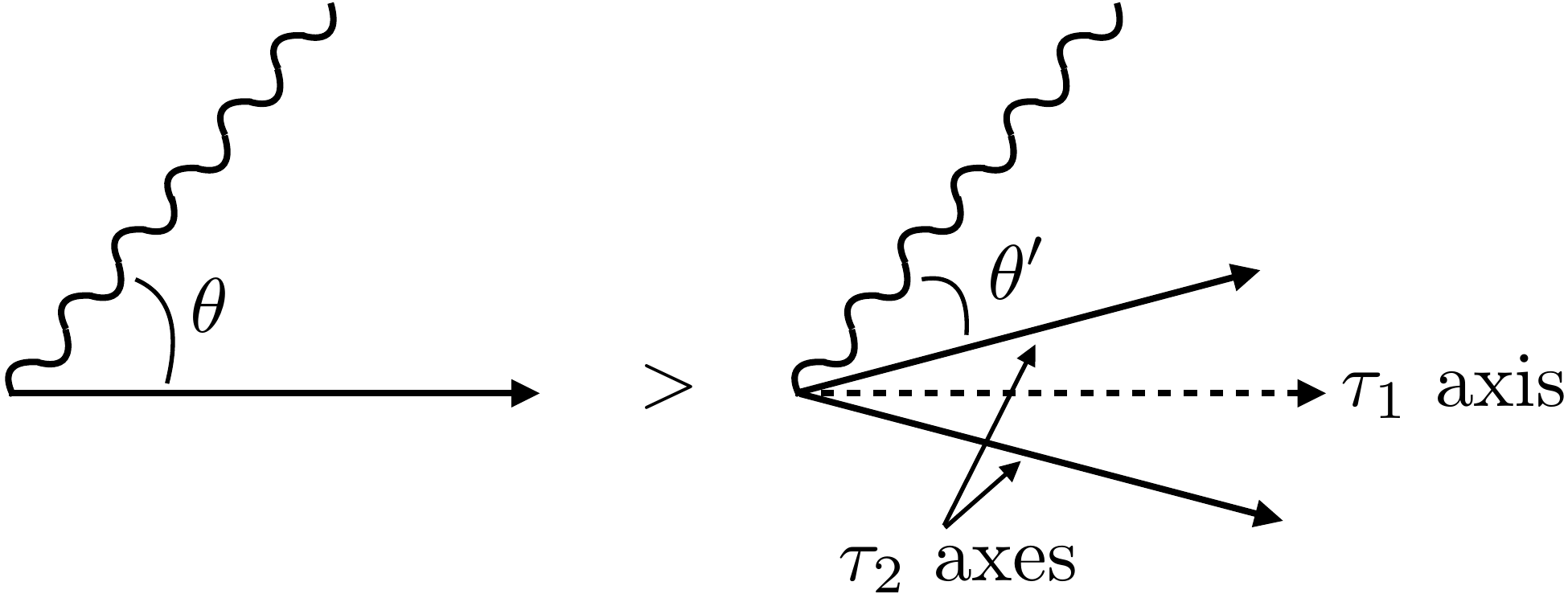}
\end{center}
Thus our phase space defined by measuring $\tau_1^{(2)}$ and $\tau_2^{(2)}$ looks like:

\begin{center}
\includegraphics[width=4cm]{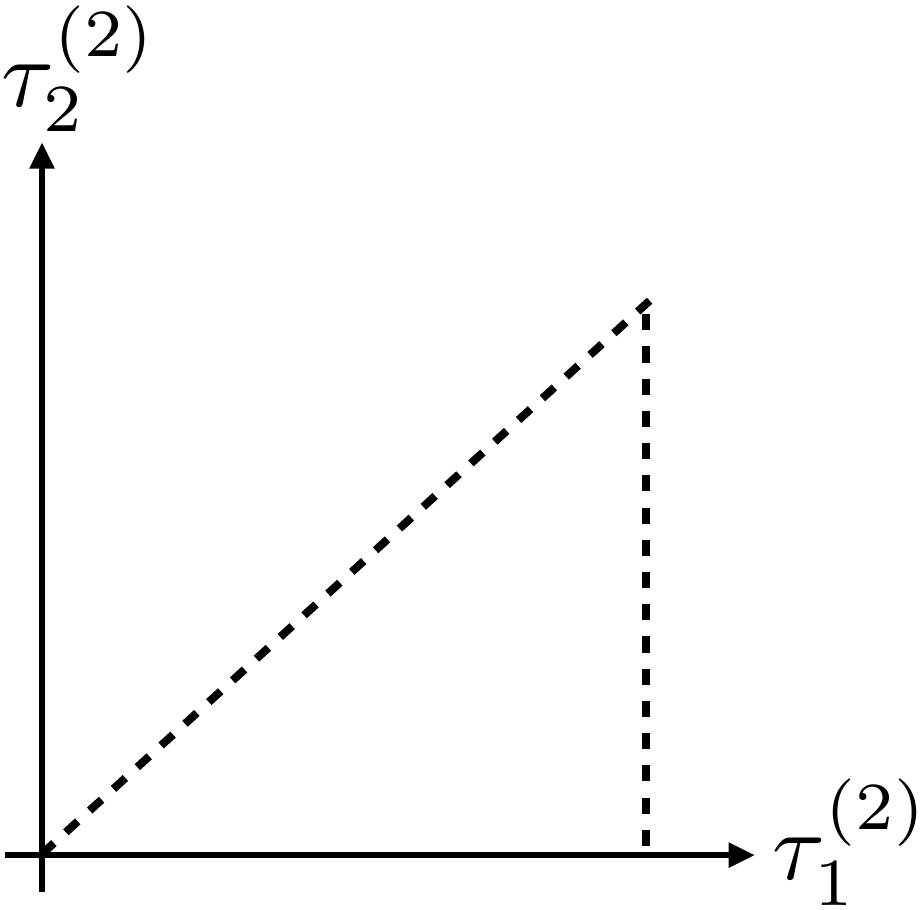}
\end{center}
Phase space is the triangular region bounded by the $\tau_1^{(2)}$ axis and the dashed lines.

To go further, we need two things: (1) identification of the soft/collinear region and (2) the form of the likelihood on this space.  (1) is easy enough to answer: because $\tau_N^{(\alpha)}>0$ and is IRC safe, the $\tau_N^{(\alpha)}\to 0$ limit is the soft/collinear divergent limit.  On our phase space, this is just the region near the $\tau_1^{(\alpha)}$ axis:

\begin{center}
\includegraphics[width=4cm]{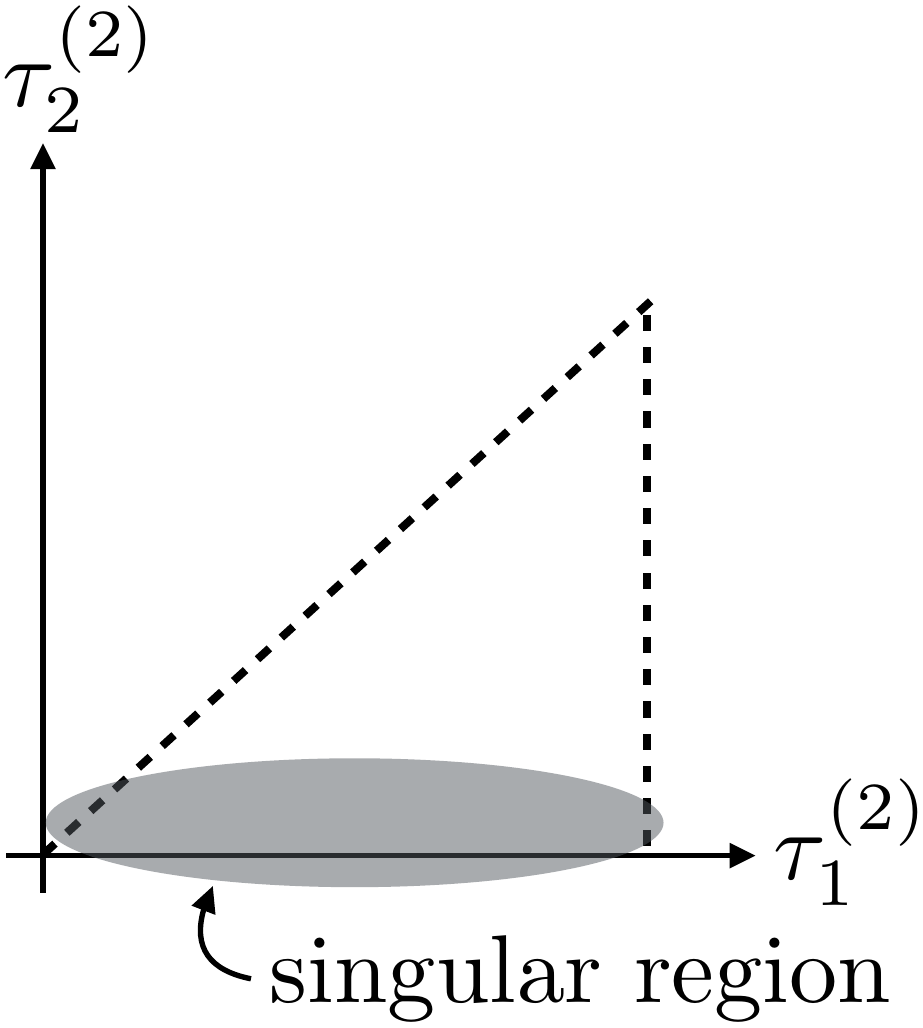}
\end{center}

Okay, let's see why this observation is so important.  The likelihood ratio of quark and gluon probability distributions on this space is
\begin{equation}
{\cal L}(\tau_1^{(\alpha)},\tau_2^{(\alpha)}) = \frac{p_g(\tau_1^{(\alpha)},\tau_2^{(\alpha)})}{p_q(\tau_1^{(\alpha)},\tau_2^{(\alpha)})}\,.
\end{equation}
Do we have any information as to what the functional form of this ratio might be?  Yes, from the Sudakovs!  By the IRC safety (and additivity) of $N$-subjettiness, the quark and gluon probabilities have Sudakov factors of the form
\begin{align}
&p_q(\tau_1^{(\alpha)},\tau_2^{(\alpha)}) \sim e^{-\alpha_s C_F f(\tau_1^{(\alpha)},\tau_2^{(\alpha)})}\,,
&p_g(\tau_1^{(\alpha)},\tau_2^{(\alpha)}) \sim e^{-\alpha_s C_A f(\tau_1^{(\alpha)},\tau_2^{(\alpha)})}\,.
\end{align}
Now, without an explicit calculation, we don't know what the functional form of $f(\tau_1^{(\alpha)},\tau_2^{(\alpha)})$ is, but, by IRC safety of $N$-subjettiness, we know its behavior in limits.  The soft/collinear limit corresponds to $\tau_2^{(2)}\to 0$ (with $\tau_1^{(2)}>\tau_2^{(2)}$), and in this limit, the Sudakov factor exponentially suppresses the probability distributions.  With this exponential form, we then must have that
\begin{equation}
f(\tau_1^{(\alpha)},\tau_2^{(\alpha)}\to0)\to\infty\,.
\end{equation}

Now, the form of the likelihood ratio for quark versus gluon jet discrimination is
\begin{equation}
{\cal L}(\tau_1^{(\alpha)},\tau_2^{(\alpha)}) = \frac{p_g(\tau_1^{(\alpha)},\tau_2^{(\alpha)})}{p_q(\tau_1^{(\alpha)},\tau_2^{(\alpha)})} \sim e^{-\alpha_s (C_A-C_F) f(\tau_1^{(\alpha)},\tau_2^{(\alpha)})}\,.
\end{equation}
Because $C_A>C_F$ in QCD, we still have that the likelihood vanishes in the singular, $\tau_2^{(2)}\to 0$ limit:
\begin{equation}
{\cal L}(\tau_1^{(\alpha)},\tau_2^{(\alpha)}\to 0)\to 0\,.
\end{equation}
However, the entire region where $\tau_2^{(2)}\to 0$ is the soft/collinear limit in which a fixed-order description of jets diverges.  In this entire singular region, the likelihood takes a unique value: ${\cal L} = 0$.  Thus, the likelihood for quark versus gluon discrimination is itself IRC safe, from our earlier discussion.  Out of all possible functions of input particle momenta, the likelihood for this problem is IRC safe, which is a strong constraint on its form.  Thus, we learn that if you want to distinguish quark-flavor from gluon-flavor jets, a good place to start is to use an IRC safe observable.  It has actually been known for a long time that IRC safe observables are good quark vs.~gluon jet discriminants \cite{Gallicchio:2011xq,Gallicchio:2012ez}.  Further, it has been demonstrated from construction of a deep neural network that the likelihood for quark vs.~gluon discrimination is consistent with IRC safety \cite{Komiske:2018cqr,Romero:2021qlf,Konar:2021zdg}.

Again, I want to emphasize that we, humans, learned something about QCD by thinking like a machine.  What else might we learn with this approach?  I hope you can find something new!

\subsection{Exercises}

\begin{enumerate}

\item Just consider the angularities for discrimination of quark vs.~gluon jets.

\begin{enumerate}

\item What is the likelihood ratio for the probability distributions $p_g(\tau_\alpha)$ and $p_q(\tau_\alpha)$?

\item What is the distribution of this likelihood ${\cal L}$ on quark jets and gluon jets, $p_q({\cal L})$ and $p_g({\cal L})$?

\item The receiver operating characteristic curve (ROC) quantifies the ``strength'' of separation power of the likelihood, as a function of a cut on the likelihood.  If the cumulative distributions of the likelihood for quarks and gluons are
\begin{align}
&\Sigma_q({\cal L}) = \int_0^{\cal L}d{\cal L}'\, p_q({\cal L}')\,,
&\Sigma_g({\cal L}) = \int_0^{\cal L}d{\cal L}'\, p_g({\cal L}')\,,
\end{align}
the ROC curve is defined to be
\begin{equation}
\text{ROC}(x) = \Sigma_g(\Sigma_q^{-1}(x))\,,
\end{equation}
where $\Sigma_q^{-1}(x)$ is the inverse of the quark's cumulative distribution function.  What is the ROC, as a function of the quark fraction $x$?

\item The area under the ROC curve (AUC) is also an interesting discrimination metric, often used by a (real) machine in a gradient descent algorithm.  What is the AUC for the ROC curve in part (c)?

\end{enumerate}

\item Consider the number of jets as defined by the procedure introduced in Exercise (2) of Lecture 2.  Consider that procedure measured on $e^+e^-\to q\bar q+X$ and $e^+e^-\to gg+X$ events, where $X$ is any other hadronic activity.  Using the discrete probability distribution $p_n$ for the quark and gluon final states, determine the likelihood ratio, ROC curve and AUC for this number of jets observable for discrimination of the quark from gluon final states, as a function of the parameter $y_\text{cut}$.  Does the AUC for this observable ever correspond to better discrimination to that for $\tau_\alpha$ from Exercise (1) above?

\end{enumerate}

\section*{Acknowledgments} 

I thank Myeonghun Park for the invitation to lecture at the QUC Winter School on Energy Frontier.



\begin{thebibliography}{1}

\bibitem{Larkoski:2017fip}
A.~J.~Larkoski,
``An Unorthodox Introduction to QCD,''
[arXiv:1709.06195 [hep-ph]].

\bibitem{Larkoski:2020jyz}
A.~J.~Larkoski,
``Another Unorthodox Introduction to QCD and now Machine Learning,''
[arXiv:2008.09673 [hep-ph]].

\bibitem{Larkoski:2019jnv}
A.~J.~Larkoski,
``Elementary Particle Physics: An Intuitive Introduction,'' Cambridge University Press (2019).

\bibitem{grifem}
D.~J.~Griffiths, {\it Introduction to Electrodynamics}, Cambridge University Press (2017).

\bibitem{Wilson:1974sk}
K.~G.~Wilson,
``Confinement of Quarks,''
Phys. Rev. D \textbf{10}, 2445-2459 (1974).

\bibitem{Kulish:1970ut}
P.~P.~Kulish and L.~D.~Faddeev,
``Asymptotic conditions and infrared divergences in quantum electrodynamics,''
Theor. Math. Phys. \textbf{4}, 745 (1970).

\bibitem{Kinoshita:1962ur} 
  T.~Kinoshita,
  ``Mass singularities of Feynman amplitudes,''
  J.\ Math.\ Phys.\  {\bf 3}, 650 (1962).
  
\bibitem{Lee:1964is} 
  T.~D.~Lee and M.~Nauenberg,
  ``Degenerate Systems and Mass Singularities,''
  Phys.\ Rev.\  {\bf 133}, B1549 (1964).

\bibitem{Bloch:1937pw}
F.~Bloch and A.~Nordsieck,
``Note on the Radiation Field of the electron,''
Phys. Rev. \textbf{52}, 54-59 (1937).

\bibitem{Andersson:1988gp} 
  B.~Andersson, G.~Gustafson, L.~Lonnblad and U.~Pettersson,
  ``Coherence Effects in Deep Inelastic Scattering,''
  Z.\ Phys.\ C {\bf 43}, 625 (1989).

\bibitem{Sjostrand:2006za} 
  T.~Sj\"ostrand, S.~Mrenna and P.~Z.~Skands,
  ``PYTHIA 6.4 Physics and Manual,''
  JHEP {\bf 0605}, 026 (2006)
  [hep-ph/0603175].
  
\bibitem{Sjostrand:2014zea} 
  T.~Sj\"ostrand {\it et al.},
  ``An Introduction to PYTHIA 8.2,''
  Comput.\ Phys.\ Commun.\  {\bf 191}, 159 (2015)
  [arXiv:1410.3012 [hep-ph]].
  
\bibitem{Bahr:2008pv} 
  M.~Bahr {\it et al.},
  ``Herwig++ Physics and Manual,''
  Eur.\ Phys.\ J.\ C {\bf 58}, 639 (2008)
  [arXiv:0803.0883 [hep-ph]].
  
\bibitem{Bellm:2015jjp} 
  J.~Bellm {\it et al.},
  ``Herwig 7.0/Herwig++ 3.0 release note,''
  Eur.\ Phys.\ J.\ C {\bf 76}, no. 4, 196 (2016)
  [arXiv:1512.01178 [hep-ph]].
  
\bibitem{Gleisberg:2008ta}
T.~Gleisberg, S.~Hoeche, F.~Krauss, M.~Schonherr, S.~Schumann, F.~Siegert and J.~Winter,
``Event generation with SHERPA 1.1,''
JHEP \textbf{02}, 007 (2009)
[arXiv:0811.4622 [hep-ph]].
  
\bibitem{Bothmann:2019yzt}
E.~Bothmann \textit{et al.} [Sherpa],
``Event Generation with Sherpa 2.2,''
SciPost Phys. \textbf{7}, no.3, 034 (2019)
[arXiv:1905.09127 [hep-ph]].

\bibitem{Berger:2003iw}
C.~F.~Berger, T.~Kucs and G.~F.~Sterman,
``Event shape / energy flow correlations,''
Phys. Rev. D \textbf{68}, 014012 (2003)
[arXiv:hep-ph/0303051 [hep-ph]].

\bibitem{Almeida:2008yp}
L.~G.~Almeida, S.~J.~Lee, G.~Perez, G.~F.~Sterman, I.~Sung and J.~Virzi,
``Substructure of high-$p_T$ Jets at the LHC,''
Phys. Rev. D \textbf{79}, 074017 (2009)
[arXiv:0807.0234 [hep-ph]].

\bibitem{Ellis:2010rwa}
S.~D.~Ellis, C.~K.~Vermilion, J.~R.~Walsh, A.~Hornig and C.~Lee,
``Jet Shapes and Jet Algorithms in SCET,''
JHEP \textbf{11}, 101 (2010)
[arXiv:1001.0014 [hep-ph]].

\bibitem{Larkoski:2014uqa}
A.~J.~Larkoski, D.~Neill and J.~Thaler,
``Jet Shapes with the Broadening Axis,''
JHEP \textbf{04}, 017 (2014)
[arXiv:1401.2158 [hep-ph]].

\bibitem{Sudakov:1954sw} 
  V.~V.~Sudakov,
  ``Vertex parts at very high-energies in quantum electrodynamics,''
  Sov.\ Phys.\ JETP {\bf 3}, 65 (1956)
  [Zh.\ Eksp.\ Teor.\ Fiz.\  {\bf 30}, 87 (1956)].

\bibitem{Larkoski:2013paa}
A.~J.~Larkoski and J.~Thaler,
``Unsafe but Calculable: Ratios of Angularities in Perturbative QCD,''
JHEP \textbf{09}, 137 (2013)
[arXiv:1307.1699 [hep-ph]].

\bibitem{Heister:2003aj}
A.~Heister \textit{et al.} [ALEPH],
``Studies of QCD at e+ e- centre-of-mass energies between 91-GeV and 209-GeV,''
Eur. Phys. J. C \textbf{35}, 457-486 (2004).

\bibitem{Ellis:1991qj}
R.~K.~Ellis, W.~J.~Stirling and B.~R.~Webber,
``QCD and collider physics,''
Camb. Monogr. Part. Phys. Nucl. Phys. Cosmol. \textbf{8}, 1-435 (1996).

\bibitem{Gross:1973id}
D.~J.~Gross and F.~Wilczek,
``Ultraviolet Behavior of Nonabelian Gauge Theories,''
Phys. Rev. Lett. \textbf{30}, 1343-1346 (1973).

\bibitem{Politzer:1973fx}
H.~D.~Politzer,
``Reliable Perturbative Results for Strong Interactions?,''
Phys. Rev. Lett. \textbf{30}, 1346-1349 (1973).

\bibitem{ParticleDataGroup:2020ssz}
P.~A.~Zyla \textit{et al.} [Particle Data Group],
``Review of Particle Physics,''
PTEP \textbf{2020}, no.8, 083C01 (2020).

\bibitem{Altarelli:1977zs}
G.~Altarelli and G.~Parisi,
``Asymptotic Freedom in Parton Language,''
Nucl. Phys. B \textbf{126}, 298-318 (1977).

\bibitem{Landau:1954cxv}
L.~D.~Landau, A.~A.~Abrikosov and I.~M.~Khalatnikov,
``An asymptotic expression for the photon Green function in quantum electrodynamics,''
Dokl. Akad. Nauk SSSR \textbf{95}, no.6, 1177-1180 (1954).

\bibitem{Larkoski:2017jix}
A.~J.~Larkoski, I.~Moult and B.~Nachman,
``Jet Substructure at the Large Hadron Collider: A Review of Recent Advances in Theory and Machine Learning,''
Phys. Rept. \textbf{841}, 1-63 (2020)
[arXiv:1709.04464 [hep-ph]].

\bibitem{Guest:2018yhq}
D.~Guest, K.~Cranmer and D.~Whiteson,
``Deep Learning and its Application to LHC Physics,''
Ann. Rev. Nucl. Part. Sci. \textbf{68}, 161-181 (2018)
[arXiv:1806.11484 [hep-ex]].

\bibitem{Radovic:2018dip}
A.~Radovic, M.~Williams, D.~Rousseau, M.~Kagan, D.~Bonacorsi, A.~Himmel, A.~Aurisano, K.~Terao and T.~Wongjirad,
``Machine learning at the energy and intensity frontiers of particle physics,''
Nature \textbf{560}, no.7716, 41-48 (2018).

\bibitem{Albertsson:2018maf}
K.~Albertsson, P.~Altoe, D.~Anderson, J.~Anderson, et al.,
``Machine Learning in High Energy Physics Community White Paper,''
J. Phys. Conf. Ser. \textbf{1085}, no.2, 022008 (2018)
[arXiv:1807.02876 [physics.comp-ph]].

\bibitem{Carleo:2019ptp}
G.~Carleo, I.~Cirac, K.~Cranmer, L.~Daudet, M.~Schuld, N.~Tishby, L.~Vogt-Maranto and L.~Zdeborov\'a,
``Machine learning and the physical sciences,''
Rev. Mod. Phys. \textbf{91}, no.4, 045002 (2019)
[arXiv:1903.10563 [physics.comp-ph]].

\bibitem{Bourilkov:2019yoi}
D.~Bourilkov,
``Machine and Deep Learning Applications in Particle Physics,''
Int. J. Mod. Phys. A \textbf{34}, no.35, 1930019 (2020)
[arXiv:1912.08245 [physics.data-an]].

\bibitem{Karagiorgi:2021ngt}
G.~Karagiorgi, G.~Kasieczka, S.~Kravitz, B.~Nachman and D.~Shih,
``Machine Learning in the Search for New Fundamental Physics,''
[arXiv:2112.03769 [hep-ph]].

\bibitem{Feickert:2021ajf}
M.~Feickert and B.~Nachman,
``A Living Review of Machine Learning for Particle Physics,''
[arXiv:2102.02770 [hep-ph]].

\bibitem{cybenko}
G.~Cybenko, 
``Approximation by superpositions of a sigmoidal function,'' 
Math. Control Signal Systems \textbf{2}, 303-314 (1989). 

\bibitem{hornik}
K.~Hornik, 
``Approximation capabilities of multilayer feedforward networks,''
Neural Networks \textbf{4} 2, 251-257 (1991).

\bibitem{leshno}
M.~Leshno, V.~Y.~Lin, A.~Pinkus and S.~Schocken,
``Multilayer feedforward networks with a nonpolynomial activation function can approximate any function,''
Neural Networks \textbf{6} 6, 861-867 (1993).

\bibitem{Neyman:1933wgr}
J.~Neyman and E.~S.~Pearson,
``On the Problem of the Most Efficient Tests of Statistical Hypotheses,''
Phil. Trans. Roy. Soc. Lond. A \textbf{231}, no.694-706, 289-337 (1933).

\bibitem{Thaler:2010tr}
J.~Thaler and K.~Van Tilburg,
``Identifying Boosted Objects with N-subjettiness,''
JHEP \textbf{03}, 015 (2011)
[arXiv:1011.2268 [hep-ph]].

\bibitem{Thaler:2011gf}
J.~Thaler and K.~Van Tilburg,
``Maximizing Boosted Top Identification by Minimizing N-subjettiness,''
JHEP \textbf{02}, 093 (2012)
[arXiv:1108.2701 [hep-ph]].

\bibitem{Datta:2017rhs}
K.~Datta and A.~Larkoski,
``How Much Information is in a Jet?,''
JHEP \textbf{06}, 073 (2017)
[arXiv:1704.08249 [hep-ph]].

\bibitem{Larkoski:2019nwj}
A.~J.~Larkoski and E.~M.~Metodiev,
``A Theory of Quark vs. Gluon Discrimination,''
JHEP \textbf{10}, 014 (2019)
[arXiv:1906.01639 [hep-ph]].

\bibitem{Gallicchio:2011xq}
J.~Gallicchio and M.~D.~Schwartz,
``Quark and Gluon Tagging at the LHC,''
Phys. Rev. Lett. \textbf{107}, 172001 (2011)
[arXiv:1106.3076 [hep-ph]].

\bibitem{Gallicchio:2012ez}
J.~Gallicchio and M.~D.~Schwartz,
``Quark and Gluon Jet Substructure,''
JHEP \textbf{04}, 090 (2013)
[arXiv:1211.7038 [hep-ph]].

\bibitem{Komiske:2018cqr}
P.~T.~Komiske, E.~M.~Metodiev and J.~Thaler,
``Energy Flow Networks: Deep Sets for Particle Jets,''
JHEP \textbf{01}, 121 (2019)
[arXiv:1810.05165 [hep-ph]].

\bibitem{Romero:2021qlf}
A.~Romero, D.~Whiteson, M.~Fenton, J.~Collado and P.~Baldi,
``Safety of Quark/Gluon Jet Classification,''
[arXiv:2103.09103 [hep-ph]].

\bibitem{Konar:2021zdg}
P.~Konar, V.~S.~Ngairangbam and M.~Spannowsky,
``Energy-weighted Message Passing: an infra-red and collinear safe graph neural network algorithm,''
[arXiv:2109.14636 [hep-ph]].

%
%
%
%
%
%
%
%
%
%
%
%
%
%
%
%
%
%
%
  
\end{thebibliography}
\end{document}